\documentstyle[11pt,epsfig]{article}
\renewcommand{\theequation}{\thesubsection.\arabic{equation}}
\begin{document}
\newcommand{\lan}{\langle}
\newcommand{\ran}{\rangle}
\newcommand{\be}{\begin{equation}}
\newcommand{\br}{\begin{eqnarray}}
\newcommand{\ee}{\end{equation}}
\newcommand{\er}{\end{eqnarray}}
\title{
\hfill\parbox{4cm}{\normalsize IMSC/2001/12/55\\
                               cond-mat/0112439}\\       
\vspace{2cm}
On the Emergence of the Microcanonical Description
from a Pure State}
\author{N. D. Hari Dass, S. Kalyana Rama and B. Sathiapalan\\
 {\em  Institute of Mathematical Sciences}\\{\em Taramani }\\
{\em Chennai, India 600113}}                                    
\maketitle                                                                 
\begin{abstract}                                                           
We study, in general terms, the process by which a pure state can
``self-thermalize'' and {\em appear} to be described by a
microcanonical density matrix.  This requires a quantum mechanical
version of the Gibbsian coarse graining that conceptually underlies
classical statistical mechanics. We introduce some extra degrees of
freedom that are necessary for this. Interaction between these degrees
and the system can be understood as a process of resonant absorption
and emission of ``soft quanta''. This intuitive picture allows one to
state a criterion for when self thermalization occurs. This paradigm
also provides a method for calculating the thermalization rate using
the usual formalism of atomic physics for calculating decay rates. We
contrast our prescription for coarse graining, which is somewhat
dynamical, with the earlier approaches that are intrinsically
kinematical.  An important motivation for this study is the black hole
information paradox.
\end{abstract}

\section{Introduction}
\subsection{Motivation}
  Understanding the principles of Statistical Mechanics from first
principles still remains a challenge [\cite{reichl}-\cite{Haake}] despite the monumental work by the
pioneers like Boltzman, Poincare etc.
  According to the fundamental principles of statistical mechanics
a macro-system starting from an arbitrary
initial state  eventually tends to a state of so called
{\it thermodynamic
equilibrium}.
Inherent in this assertion is a certain 
{\em irreversibility}
and one of the main issues has been the reconciliation
of this with the fact that basic dynamics is 
{\em reversible}.
In the case of Quantum Statistical Mechanics there are
additional complications
as the thermodynamic equilibrium state, which
 is described by a {\it mixed density matrix}
can never be obtained from a  {\it pure}
density matrix under quantum mechanical evolution which
is described by unitary transformations.
This last issue is also the crux of the so called
{\em Quantum Measurement Problem}.
Interest in these issues has recently been rekindled
from a most unexpected direction i.e the problems
of black hole entropy and of the so called black hole
information. In this section we briefly describe each
of these to bring into focus the issues we address in the rest of the paper

\subsection{Foundations of Quantum Statistical Mechanics}

The questions discussed above lead naturally to questions about the
basic postulate of quantum
statistical mechanics(QSM). The postulate is that of equal a priori probabilities for all
microstates in a given macrostate of definite energy. Equivalently,
the density matrix
of an ensemble whose energy lies in the range $E_0$ to $E_0+\Delta E$ is given
by the identity
matrix in the energy basis, suitably normalized as shown
($N$ is the number of states in the given energy interval):

\be   \label{1.1}
\rho = {1\over N}I  \; \; \; E_0 \le E \le E_0 + \Delta E
\ee
and for $E$ not in this interval,
\[
\rho =0
\]
It is worth pointing out here that the standard
derivation of the microcanonical distribution 
can not be carried over automatically to QSM.
In classical statistical mechanics one derives the
microcanonical distribution, by maximizing the entropy:
\be
S = -k_B\int dX^N \rho(X^N) log \rho(X^N)
\ee
subject to
\be
1= \int dX^N \rho(X^N)
\ee
In these eqns $\rho(X^N)$ is the {\em distribution 
function} on the {\em phase space}.
The result one obtains is
\be \rho(X^N) = const ~~~~~ E_0\le H\le E_0+\Delta E
\ee
and zero otherwise.
It is often stated that  for quantum systems the proof
is similar with the density operator replacing
$\rho$, the quantum mechanical density matrix
operator \cite{reichl}:
\be
S = -k_B tr \rho \log\rho~=-k_B\sum_n P_n\log P_n
\ee
where $P_n = \lan E,n|\rho|E,n\ran$(the fact that $\rho$
is a constant of motion has already
been used).
But for a pure state the eigenvalues of $\rho$ are $0,1$
(only one eigenvalue 1
and all
others 0). Hence the entropy is exactly {\it zero} 
and no variational principle is applicable.\\

It is sometimes argued that what one deals in practice
is not $\rho$ itself but the time-averaged $\rho$:
\be
\bar\rho(t) = {1\over 2T}\int_{t-T}^{t+T}~dt^{'} ~\rho(t^{'})
\ee
It follows trivially that
\be
 tr~\bar\rho = 1
\ee
But $tr {\bar\rho}^2\ne 1$  as can be easily shown using:
 $\rho(t) = e^{iHt}~~\rho(0)~~e^{-iHt}$:
\be
{\bar\rho(t)}^2 = {1\over 4T^2}\int~dt^{'}~dt^{''}~\rho(t^{'})~ \rho(t^{''})
\ee
Though $tr~{\bar\rho}^2~\ne 1$ and therefore the time 
averaged ${\bar\rho}$ represents a
mixed state,
$tr~{\bar\rho}^2$ is calculable in terms of the initial density matrix:
\be
tr~~{\bar\rho(t)}^2~~= \sum_{n,m}~~|\lan n|\rho(0)|m\ran |^2({\sin(\Delta E_{nm} ~~T)\over
(\Delta E_{nm}~~T)})^2
\ee
In fact traces of all powers of ${\bar\rho}$ are calculable
and found to be completely determined by the initial
pure density matrix and the spectrum of the Hamiltonian
describing the system. The important point is that all
these constraints have to be used in the variational
method and subsequently one does not get the
microcanonical distribution.

One can ask, what is the nature of a quantum mechanical
system that has the property that, even though we start
it off in a pure state, it evolves after some suitable
time $\tau$ to a state that {\em for all practical 
purposes} (which means for all observables of interest) gives the
{\em same expectation} values as a system described 
by the
microcanonical density matrix.
To mimic the situation in classical statistical mechanics
where we consider an initial configuration with energy
lying in a narrow band,
it is assumed that the pure state is a linear combination
of energy eigenstates with energy between $E_0$ and
$E_0+\Delta E$. We should also discuss what happens when
the pure state itself happens to be an energy 
eigenstate. We will
give an answer stated in terms of the properties of
the exact eigenfunctions of the system. 
This will be the main result of this paper. 
It seems to us that there
is an inbuilt dependence on the choice of
``interesting'' observables which define the coarse-grained
microstates.
\subsection{Quantum Measurement Problem}
According to the von Neumann {\it projection postulate}, also
known as the '{\it collapse of the wave-function}' 
postulate,
if an observable $A$ is measured in a generic quantum 
state,
the result will be any one of the eigenvalues of $A$ 
and the
state after the measurement {\it collapses} to the 
corresponding
eigenstate. This implies a transition from the initial
{\it pure ensemble} characterized by
\be
 tr~ \rho = 1, tr~ \rho^2 = 1
\ee
before the measurement to a {\it mixed ensemble}
with
\be
tr~ \rho = 1, tr~ \rho^2\ne 1
\ee
Let us illustrate this with a simple example. Consider 
an ensemble of states $|\psi>$ of a two-level system 
on which a measurement is done whose possible outcomes 
are $s=\pm$ with the corresponding eigenstates 
$|+>,|->$.
As the latter span a basis for the two-level Hilbert 
space
we could expand $|\psi>$ in this basis
\be
|\psi>~=~c_+|+>~+c_-|->
\ee
The initial density matrix
\be
\rho_i=|\psi><\psi|
\ee
is {\em pure}.
After the measurement we have a {\it mixture} of two 
pure ensembles of states $|+>,|->$ with weight 
factors $|c_+|^2, |c_-|^2$ resulting in the 
final density matrix
\be
\rho_f= |c_+|^2~|+><+|~+|c_-|^2~|-><-|
\ee
It is easy to check that
\be
tr~ \rho_f = 1, tr~ \rho_f^2=(|c_+|^2)^2+(|c_-|^2)^2
= 1-2|c_+|^2|c_-|^2\leq 1
\ee

As  {Unitary} transformations preserve all traces
\be
\rho^{'} = U\rho U^{\dagger}\rightarrow tr~ {\rho^{'}}^n
= tr~ {\rho}^n
\ee
the measurement process is not describable by a unitary
transformation and is thus {\it irreversible}. It is as
if the measurement process has to be treated 
differently from ordinary dynamics. This is the crux 
of the so called {\em Quantum Measurement Problem} 
and one sees that once again the crucial
issue has arisen of pure states evolving into 
mixed states,
or more plausibly into pure states that {\em for all 
practical purposes} look like mixed states.
\subsection{Black hole evolution and AdS/CFT 
correspondence}
One of our strong motivation for this study comes 
from an issue related to
 the ``information paradox'' in black hole physics.
The issue is the following: Consider the quantum 
mechanical
description of black hole
formation. Some matter/energy in some initial state
described by a wave function evolves in time and at 
some point
makes a transition to a black hole state due to 
the attractive gravitational interaction. As per the
Hawking effect the black hole 
appears to be radiating
like a blackbody at its Hawking temperature. The 
system does not look to an external observer to be 
described by a pure state wave function but rather by 
a mixed state density matrix characterstic of a thermal
state. The usual argument is that this transition
from a pure state to a mixed state is {\it illusory} 
because when we include the degrees of freedom 
{\it inside} the black hole one recovers a pure state
description. Indeed string theory has given us a 
prescription in some situations for actually
counting the number
of microscopic states associated with a black hole 
which reproduces the Bekenstein entropy. Nevertheless,
(and this is the crucial point), while ignoring the 
degrees of freedom in the
interior of the
black hole can make the pure state appear mixed to 
the external observer,
this does not explain
why it should look {\it thermal}. In other words, 
when calculating the entropy
by counting the
number of states, there is an {\em implicit} 
assumption that the interior system
is a {\em mixture} of states with {\em equal} a priori
probabilities i.e it is ergodic and is described by a 
microcanonical ensemble. In
ordinary statistical systems
there is always a ``heat bath'' one usually invokes - 
basically the environment - that will ensure
this, but in the case of the black
hole we are describing a closed system. There is no 
environment or heat bath.
So the question is
how does one justify such an assumption of ergodicity 
for a closed system?

In \cite{KRBS} this problem was approached using the 
so called AdS/CFT correspondence \cite{maldacena,witten}. Using this
correspondence the gravitational problem is mapped 
to a Yang-Mills problem. How does a pure state with
some fixed energy in a Yang-Mills theory 
``self-thermalize''? The solution proposed
was that chaos
would do the job. Classical Yang-Mills has already been
shown to be chaotic
\cite{Biro,Muel}. One
expects chaos
to develop and make the system ergodic. Assuming this 
is true quantum mechanically also,
this phenomenon could be mapped back to the black-hole via the same AdS/CFT correspondence.

While not much is known theoretically about quantum 
chaos in Yang-Mills theories there is
some experimental
evidence in heavy ion collisions for the formation of 
a quark-gluon plasma (at finite temperature).  The 
initial state, which is two heavy nuclei traveling
towards each other, is definitely described by a 
quantum mechanical wave function.
The
final state appears to be describable by a system
at finite temperature. If this happens, then it is 
quantum mechanical
``self thermalization'' - a pure state evolves 
unitarily by Hamiltonian evolution
and after
a while {\em looks like} a thermal state.

\subsection{Criterion for a Physical System to Become 
Ergodic.}

Given that we are able to characterize the nature of 
eigenfunctions for a system that
is ergodic, we can ask when does a physical system satisfy these properties ?
i.e. when
do we expect thermalization? 
We will give an approximate answer to this question
 in this paper.
The centre of mass degrees of freedom are expected to look thermal
most of the times. The interesting question concerns the {\em internal}
degrees of freedom that are normally frozen to some discrete quantum states.
In the Yang-Mills example above, we expect
thermalization of quark and gluon degrees of freedom to
take place only when the energy density exceeds a critical density. The scale is
obviously set by $\Lambda _{QCD}$. Similarly in the black hole example
 we do not expect
every (zero temperature) neutron star to form a black hole and produce a non-zero
Hawking temperature. This is discussed further in sec. 2.6.4.

\subsection{Thermalization Time}

If a physical system is expected to become ergodic, 
we can ask what is the time scale over which this 
happens. This is also the time scale for return to 
equilibrium when the system is disturbed. It should of
course be emphasised that depending on how the system
is disturbed, there could be many such time scales. We 
will give approximate answers to such questions.

\subsection{Non-Quantum-Mechanical and 
Quantum-Mechanical Coarse Graining}

There have been several attempts to obtain effective 
thermal density matrices, or, to obtain irreversibility
from reversibility. All these involve (as they must 
indeed) some coarse graining. However, to our knowledge
many of these involve ad hoc prescriptions that do not
arise {\em naturally} in quantum theory.  For example, 
some invoke averaging over the time of measurement. The
argument is that every measurement takes a finite
time. Nevertheless averaging over time alone is not 
sufficient.
As described in Section 1.2 it also does not give
the right answer.
Similarly, some invoke
averaging over an ensemble of initial states. 
In quantum mechanics there is no need for any other
ensemble than that required by the probabilistic
interpretation of the quantum state. Thus averaging
over some distribution of initial states is generically
ad hoc and unwarranted. The point is that in 
generic situations there is no justification for such 
ad hoc averaging procedures.
Thus whatever
coarse graining is necessary must arise naturally 
within the framework of quantum mechanics.
We will see that invoking some unobserved
``soft'' (low energy) degrees of freedom can naturally
accomplish the required coarse graining. An example 
of such degrees of freedom is
the ``soft'' photons of QED.

\subsection{Outline}
In Section 2 we will describe our proposal for Gibbsian
coarse graining in
quantum mechanics using
soft quanta. This will
address the criticisms of section 1.7. We will also 
give a physical
criterion for thermalisation-
a qualitative answer to the question of Section 1.5.
In Section 3 we introduce the mathematical formulation 
of the problem.
We also give a brief description of the work of 
von Neumann
and van Kampen on this subject (about which we learnt 
well after completing our work ). This is included 
mainly for completeness and comparison,
and is not logically required for understanding 
our work. In Section 4
we  discuss the
issues raised in Section 1.2 regarding the 
characteristics of a system that
can be described by quantum statistical mechanics.  
In Section 5
we look at the same question from a different 
perspective - that of
soft quanta and resonant transitions.
In Section 6 we will discuss a two level system
coupled to a continuum of soft quanta states. This will illustrate some of the ideas
more quantitatively. By using the quantum mechanical 
formalism underlying the familiar ``Fermi Golden
Rule'', we will see approximate irreversibility coming 
out of a reversible dynamics
and the approximate emergence of a thermal 
microcanonical density matrix.
More quantitative answers to the questions of 
Section 1.5,1.6 and
1.7 will be given here.
Section 7 will summarize the results of this paper.

\section{Gibbsian Coarse Graining and the Importance of Soft Modes}
\setcounter{equation}{00}

\subsection{Classical Gibbsian Coarse Graining}

An important ingredient in classical statistical mechanics is the notion
of coarse graining. As was originally argued by 
Gibbs \cite{gibbs}, unless the microstates
are coarse grained, entropy will always remain 
constant. This is easy to see
- as the region of phase space( energy surface) 
that is occupied by the ensemble (call this
region $\Gamma^*$; care should be taken to distinguish
$\Gamma^*$ from $\Gamma$ which usually denotes the
{\em entire} energy surface) spreads
out in phase space, its shape becomes very complicated 
but its
volume remains fixed. 
The  entropy  is given by $S= \sum _i p_i ln \; p_i$ 
where $i$ labels the microstate
and $p_i$ is the probability that the system is 
in this state.
If microstates are taken to be 
points, then 
$p_i=1$ if the point $i$ belongs to $\Gamma^*$ and zero
otherwise. It is easy to see that this number is 
constant because it depends only on the
volume of $\Gamma^*$, not the shape. If, on the other 
hand, the microstates are taken to be
small boxes(but large enough to have many system points
within them) in phase space, then $p_i$ will be equal 
to the fraction of the
box that is inside $\Gamma^*$. In this case, it is clear that as $\Gamma^*$  
spreads, most of the $p_i$'s will become equal to each other and approach
a value between 0 and 1. Thus $S$ will increase till it reaches a maximum.

\subsection{Quantum Mechanical Gibbsian Coarse Graining- ``Soft Quanta''}

What we need is the quantum mechanical analogue  of this coarse graining.
It is  important that one should stay within the formalism of quantum
mechanics while describing this coarse graining. We propose the following scheme:
Let us consider our Hilbert space to be a {\em tensor}
product of two Hilbert spaces ${\cal H}_i$ and 
${\cal H}_a$ i.e ${\cal H}={\cal H}_i\otimes{\cal H}_a$.
States in ${\cal H}$ are the {\em microstates} of
our system. $i$ are the degrees of freedom that one is 
physically
interested in and will represent the "coarse grained"
microstates. $a$ represents some degrees of freedom that we are not
interested in and possibly over which we have no control.
These are the degrees that allow us to coarse grain. As an example
consider a gas of molecules. $i$ could be the usual microscopic degrees
of freedom that one
associates with the gas, the positions and momenta of the molecules. One
can also include rotational or vibrational degrees if one wants. $a$
can be soft photons. i.e. the molecules interact with the electro-magnetic
field all the time and constantly emit and absorb radiation. There are
very long wavelength photons of almost zero energy that one has no control
over. They constitute a continuum of gap-less 
excitations in this system.
They interact with the molecules but take away negligible energy. The energy
hyper-surface defining the microcanonical ensemble in quantum mechanics
always has a small but finite width.
We can understand this as follows: We can specify the variables $i$.
The $a$ variables are not in our control. We can think of $i$ as labeling
an energy eigenstate of the $i$ system when there is no interaction with the
$a$ system.  When we turn on interactions with $a$, the state $i$ is no longer
an energy eigenstate. It becomes a linear combination of states with energy
in a range
$\Delta E$. This is an estimate of $\Delta E$ in (\ref{1.1}).
The observables of interest $O$ will be assumed to be
functions only of $i$ and not of $a$. To be precise we assume that
\be \label{CG}
\langle i,a | O | j,b \rangle = \delta _{ab}\langle i |O|j \rangle
\ee

Thus our coarse graining will be defined by saying that these are the
operators of interest and it is with respect to these operators that
 the system looks thermal. Note that this is different from the following
kind of coarse graining: the observable $O$ depends on both $i$ and $a$, i.e.,
 $\lan  i,a| O | j,b \ran $ is not necessarily given by \ref{CG}.
 Nevertheless, the $a$ variables are not in our control. So we average
 over them in some fashion - for e.g. $ {\bar O _{ij}} = \int da db
 \lan i,a | O | j,b \ran p_{a,b}$. This defines a coarse graining. But
 this is ad hoc and definitely is not
 a quantum mechanical operation. We do not want this type of coarse graining.

There is some arbitrariness in what we call $i$ and what we call $a$. But
it is crucial that the $a$ degrees are gap-less, or at least the gap
$\delta _a$ between two consecutive energy levels of the
unperturbed $a$ system should be much smaller than
the inverse of the time interval during which we observe the system.
Thus for instance, if $\delta _i =0$ for some of the 
$i$-states, then some of the $i$ degrees of freedom
could be called $a$.
Thus if $i$ represents a continuum of harmonic oscillators, then some subset
of these around zero frequency can be called the $a$ variables. Since
our energy resolution is always finite, these are not observables of interest.

\subsection{Paradigm of Discrete States Coupled to a Continuum}

 If  it is the case that we can describe the exact system as
a coupled $i$-$a$ system with $i$ being discrete and $a$ being continuous, 
then
we are familiar with this in atomic physics situations. In this situation if
we focus on the discrete system alone it will appear to be described by
a non-Hermitian Hamiltonian with complex energy eigenvalues. The width
$\Gamma$ represents the finite lifetime of the (excited) states. This
 requires
that we take the limit $\delta _a \rightarrow 0$. If we keep $\delta _a$
finite then we expect that after a finite time of O(${1\over \delta_a}$)
the system will look periodic. Thus the apparent irreversibility is only
because the time of observation is short compared to $1\over \delta _a$.
Conversely if we are to observe an irreversibility, as in the second law
of thermodynamics, it is clear that
 we need such soft degrees of freedom.  ``Soft'' in this context
means that the energy of these quanta should be less than
the bandwidth, $\Delta E$,
that defines the microcanonical system as :
$E_0 - \Delta E \leq E \leq E_0 +\Delta E$.
Thus we want $\hbar \omega < \Delta E$.
In all the
usual physical systems there are always soft quanta. So this is not an
issue.

\subsection{Heuristic Criterion for Ergodic Density Matrix}

However the above condition is not sufficient. Arguments in the next section
suggest that
matrix elements between $|i \rangle$ and $|j\rangle$ induced by the
interactions with $a$ should be much larger than $\delta _i$. This ensures that
an exact energy eigenstate contains a large number of different $i$ states.
This is not
always satisfied. Systems that do not satisfy this will not be ergodic.
In a
quantum system with an energy gap (particles in a box) one does not expect
any kind of ergodicity when the total energy of the
system is such that only the lowest
energy levels
are excited. \footnote{We remind the reader that we are discussing
closed systems and there is no heat bath.}

\subsection{Resonance and Soft Quanta}

When the energy of the soft quanta is equal to $\delta _i$, we
expect resonant absorption/emission to take place.  This induces the large
off-diagonal matrix elements between $|i\ran $ and $|j\ran$ states that
we referred to above.  Because of resonance, the coupling between the soft
quanta and the $i$ system need not be large for this to happen.
Furthermore we expect
both states $|i\ran$ and $|j\ran$ to be equally populated if the
probability, $P_{ij}$
of $|i\ran \rightarrow |j\ran$ is equal to
$P_{ji}$. This will be the case when the induced emission dominates the
spontaneous emission, which means that there should be a large number
of soft quanta available, i.e. the energy/degree of freedom in the $a$-system
should be much larger then the largest of the energy gaps $\delta _i$
which is $\Delta E$. 
Thus the criterion for thermalization is:

i)there should be a {\em continuum} of soft quanta
with energies in the range
$\delta _i$ to $\Delta E$. This ensures resonant transitions.

ii)The number of soft quanta in each mode should be $>>1$. This ensures the equality
of the upward and downward transitions and hence equality of
occupation probability.

In section 6 we will study
a quantum system where some calculations can be done perturbatively. The
results support the general picture.

\subsection{Examples:}

\subsubsection{$H_2$-molecules}

Let us turn now to some physical examples to illustrate these ideas.
Consider a gas of hydrogen molecules.
Clearly, as far as the centre of mass
degrees of freedom of the molecules are concerned the system is 
likely to be ergodic. This is again a situation where $\delta _i =0$.
When $\delta _i=0$ the off-diagonal elements of the full
Hamiltonian are clearly larger than $\delta _i$ and one expects ergodic
behaviour by the above arguments. The centre of mass motion
is in any case ergodic.

Let us focus our attention on the relative coordinates of the atoms
in one molecule and consider this as the $i$ system.
We would like to understand whether 
``self thermalization'' can take place for this sub-system.
We could alternatively
consider all the molecules, but the interaction between the internal
degrees of freedom being negligible, this will just be many copies
of the one molecule system.  
 The
$a$ states are a subset of the degrees of freedom
of the rest of the gas molecules that interact with the internal degrees
of this molecule. We will coarse grain over these degrees
by considering observables that  depend only on $i$. We
assume that the energy exchange
between the internal degrees of this molecule and that of the rest of the gas
molecules (which includes the `$a$` degrees) is very small -
smaller than the resolution of our
experiment. It therefore makes sense to talk of the microcanonical density
matrix for the $i$-system.
Let us assume that the kinetic energies
of the molecules are small and all center of mass
degrees can be included in the $a$-system. \footnote{In a
realistic system it may be that these conditions are satisfied at
such low ``temperatures'' that the $H_2$-gas may have liquefied. But these
considerations can apply for any state of matter. }

We assume that the potential energy and the energy 
levels of this system
are as shown in figure 1. The width of the dashed line
represents the energy resolution $\Delta E$ that defines a microcanonical
system.  The energy of the soft `$a$` quanta has to be less than $\Delta E$

\begin{figure}[htbp]
\begin{center}
\epsfig{file=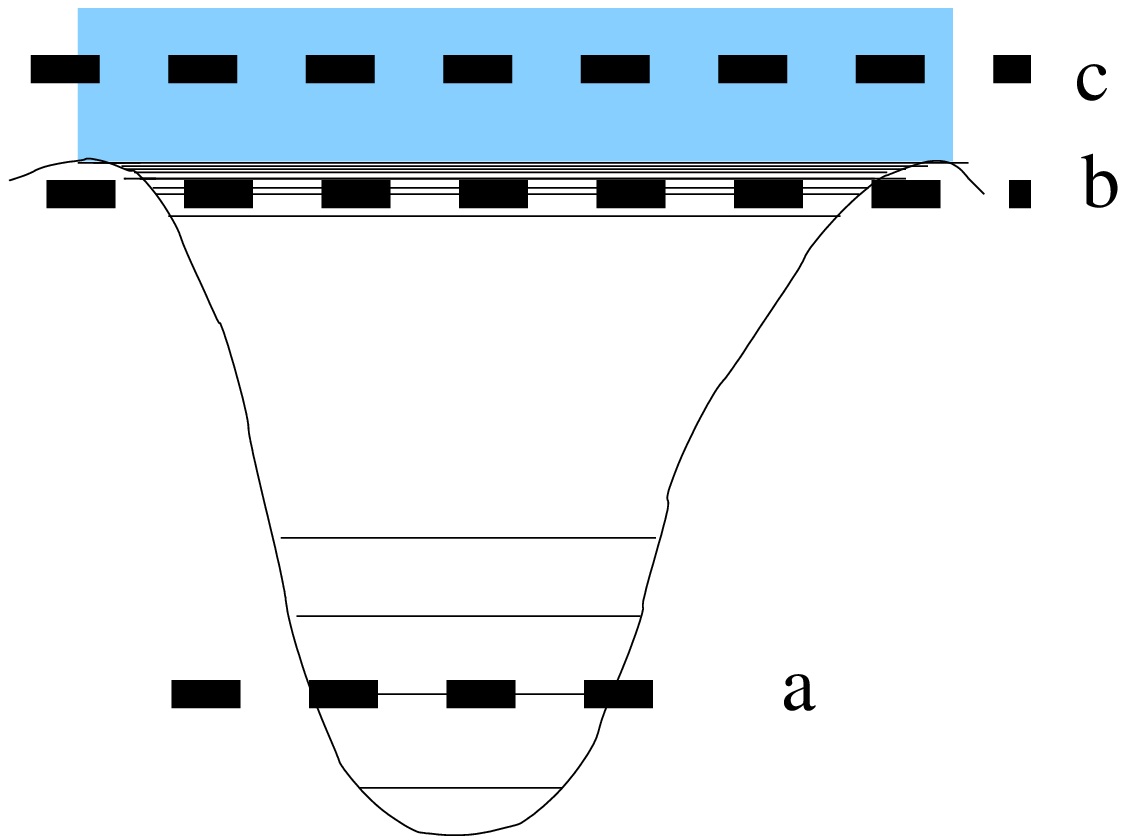, width= 12 cm,angle=0}
\vspace{ .2 in }
\begin{caption}
{In the energy range ``a'' the system is unlikely to be
ergodic. In ``c'' it is very likely ergodic,
and in ``b'' it is almost ergodic. The width of the dashed
line represents $\Delta E$   }
\end{caption}
\end{center}
\label{fig1}
\end{figure}

If the energy of the system described by the Hamiltonian for relative
motion is in the region shown by the dashed line marked ``a'',
where $\delta _i$ is fairly large compared to $\Delta E$ we do
not expect ergodic behaviour. We expect that if we start the molecule off
in a pure state, such as the ground state or first excited state,
it will remain there. This follows from purely energetic considerations.

The region marked ``c'' is a continuum
where the molecule has dissociated and we have atomic Hydrogen. In this
region $\delta _i \approx 0 << \Delta E$ and one expects ergodicity.

Region marked ``b'' is in between, with $\delta _i \approx \Delta E$
and one can expect ergodicity if
the off diagonal matrix elements are large compared to $\delta _i$. As long as
there are soft quanta that can be absorbed or emitted whose energies
match the energy spacings ($\delta _i$), this will be the case. The soft quanta
are provided by the inelastic collisions with other molecules.
We thus expect that even if we start
the system off in a pure state it will soon make transitions to the
numerous other states in that energy band and be effectively
described by a microcanonical
density matrix. This is what we would like to demonstrate in Sections 4,5 and 6.
 
As the above example illustrates, the centre of mass degrees
in a gas, for instance, typically
are expected to be ergodic. They correspond to the $\delta _i =0$
case. Classical analysis of ``billiard balls'' in certain
situation shows ergodicity
even with a small number of balls \cite{Sinai}. We need not expect
the quantum behaviour of this weakly interacting system to be very different.

The difficult question  concerns  the non-center-of-mass degrees that have
non-zero $\delta _i$. Are they ergodic or not?
 It can happen sometimes that
 while the center-of-mass degrees are ergodic, the
 non-center-of-mass degrees are in a pure state. The point we are
trying to make is that the value of the energy spacing $\delta _i$,
is a crucial
factor in deciding the answer.

\subsubsection{QCD Plasma}

Another example is that of heavy ion collision producing quark gluon plasma 
Clearly when the energy of collision is smaller than the energy gap
in QCD the hadrons are still usefully described by wave functions.
Once this gap is exceeded one can expect thermalisation. What are the soft
quanta here? Presumably these are soft gluons. There are some points of
similarity between this system and the hydrogen molecule example discussed
above. Deconfinement here, presumably corresponds to the dissociation of the
molecule there.

\subsubsection{Sodium Vapour}

Let us take another example, where again we expect to see two different
kinds of ergodicity. As it turns out this example has 
some similarity with the black hole
situation to be discussed later. Consider a vapour of sodium atoms. The centre of mass motion
of the atoms is ``gap-less'' and can be expected to be ergodic motion,
even at low temperatures. The electrons
in the atom however are described by pure state wave functions, at temperatures
that are low compared to electronic level spacings. Consider what happens when
the vapour cools down sufficiently that it forms a Sodium crystal, at some low
temperature. The centre of mass motion is no longer ergodic. However sodium
being a metal, the outer electrons are free to move around in the crystal.  
They
are now described by a Fermi gas and their motion is ergodic!
One can presumably associate a temperature with them.
The gap $\delta _i$ between electronic levels has changed from, say, a
few electron volts to something of the order $1\over L$ where $L$ is the
macroscopic size of the crystal. This is a quantum mechanical effect. Thus 
a phase transition in the material has introduced new gap-less excitations
that demonstrate ergodic behaviour, even though at a higher temperature
they were not ergodic (because they were confined to the interior of
the sodium atom)!

\subsubsection{Stars and Black Holes}

There is a similarity between the sodium vapour example and the black hole example.
When matter collapses to form a star the centre of mass motion of the
lumps of matter once again can be assumed
to be ergodic and as a consequence we get a hot star.
(Indeed astrophysicists routinely treat this motion entirely classically
using classical statistical mechanics, until we get to high densities.)
At some point the star (if sufficiently massive)
cools to essentially zero temperature, undergoes
a phase transition and forms a black hole. At this point some other
internal degrees of freedom (D-branes in string theory black holes)
 become (presumably gap-less and) ergodic
and there is a temperature that can be associated with them. These are
like the outer electrons of the sodium atom.

\subsection{Summary}

Let us summarize the points made in this section:

1. Coarse graining is done by introducing soft quanta $a$ in addition to
the conventional microscopic degrees $i$. These soft modes induce transitions
between the $|i\rangle$. The gap $\delta _a $ must be very small compared
to the inverse time of observation. Also
the off diagonal terms in the Hamiltonian connecting different $i$ states
must be much larger
than the unperturbed gap $\delta _i$. This is necessary for the exact energy
eigenstates to satisfy the properties assumed in Section 3.
This is thus necessary for ergodicity. In the resonance picture this will automatically
be the case if there is a sufficient number of soft quanta with a continuum
of frequencies in the
appropriate energy range.

2. The interaction with the continuum $a$ degrees of freedom,
induces an anti-Hermitean part to the
effective Hamiltonian of the $i$ system. This makes the evolution {\em look}
irreversible. The apparent lack of reversibility has
to do with treating the $a$-degrees as 
{\em continuous}. If a gap $\delta_a$ is introduced,
the recurrence time is $O({1\over\delta_a})$.

In section 6 
we will calculate, 
in a simplified model,
the time evolution
of the partially traced density matrix that describes the $i$ system and show
how it appears to evolve in time to a thermal one.

\section{Mathematical Formulation of the Problem and Earlier Approaches}
\setcounter{equation}{00}

In this section we give a mathematical formulation
of the problem we wish to address in this paper. It
is helpful to start with a brief recapitulation of the
situation in classical statistical mechanics. In
particular we shall follow the Gibbsian route formulated
on the classical phase space and even there we will
be concerned only with the microcanonical ensemble
description.

In such a situation one considers an isolated system
with its total energy in a narrow band
$E_0-\Delta<E<E_0+\Delta$. Under time evolution governed
by the dynamical equations the system moves on
this hyper-surface of the $6N$-dimensional phase space
called the "energy surface". Different regions of this
energy surface can be labeled by values of observables
other than the Hamiltonian. Generically these
observables are time-dependent. Further, the accuracies
with which these observables can be determined make it
meaningful to decompose the energy surface into
elementary cells called "phase cells" and all system
points within the same cell are ascribed the same value
for the "coarse grained" observables
Due to dynamics the system point moves from
phase space cell to phase space cell.The crux of the
Gibbsian statistical mechanics is the so called
ergodic theorem( more precisely the quasi-ergodic
theorem) which states that after a sufficiently long
time the system point passes through all the phase
cells and furthermore the time spent in each phase cell
is proportional to its volume. This immediately allows
one to equate the time average of observables with
an "ensemble average" wherein the weight factor for
each phase cell is its volume. As the phase cells
can be constructed with equal volume the ensemble
average can be taken with a distribution giving
equal weights to the different phase cells. This is the
microcanonical distribution. The crucial point is that
the time average has been replaced by an ensemble
average in such a way that the distribution is
{\em independent} of the initial state. It should be
emphasized that in addition to the quasi-ergodic
theorem one also approaches the same problem through
the so called H-theorem and the concomitant concept
of entropy.In what follows we shall only concentrate
on the quasi-ergodic aspects of the problem.

Having stated all this it is very important to
emphasize that a precise proof of the quasi-ergodic
theorem and in particular the resolution of the precise
role of dynamics is a very difficult problem. Both in
this formulation as well as in the Boltzmannian approach
to statistical mechanics assumptions have to be made
that are equivalent to the assumption of chaotic
behaviour.
\subsection{Formulation of Quantum Statistical Mechanics.}
Now the main question is how to formulate and prove
similar statements in the context of quantum mechanics.
The foremost difficulty here is that unlike as in
classical mechanics there is no concept of a phase
space nor of a trajectory.If classically one views
the phase space as the space of all possible states
of the classical system, the natural analog in the
quantum case is the Hilbert space. Already at this
stage many crucial differences appear; one such is
the fact that for even a simple system like the
ideal gas while the classical phase space is
{\em finite-dimensional} with dimension $6N$, the
quantum mechanical Hilbert space ${\cal H}$ is
the tensor product of $N$ copies of Hilbert spaces
${\cal H}_i$ each of which is {\em infinite-dimensional}.

Let $H$ be the {\it exact} Hamiltonian of the system and let $|A\ran$
be the {\it exact} eigenstate with eigenvalue $E_A$. Now Consider energy
eigenstates such that their eigenvalues are in the range
$E_0-\Delta<E_A<E_0+\Delta$. Further, let the initial state of the system
$|\psi\ran$ be such that
\be \label{1}
|\psi\ran = \sum_A C_A^\psi |A\ran
\ee
Then
\be \label{2}
|\psi(t)\ran = \sum_A~C_A^\psi~e^{-iE_At}~|A\ran
\ee
If we denote $C_A^\psi(t)=C_A^\psi e^{-iE_At}$ it
is obvious that $|C_A(t)|^2$ are {\em independent}
of time {\em for all A}. The average energy of the
system at any time t is given by
\be \label{3}
\bar E(t) = \sum_A |C_A(t)|^2 E_A
\ee
>From eqn (\ref{3}) we see another principal difference
between the classical and quantum situations: in the
quantum case not only is $\bar E$ constant in time, but
so are all the quantities $|C_A|^2$. If we now decompose
$C_A(t)$ into its magnitude $r_A$ and its phase $\theta_A$
 we find that while $r_A$ does not change with time,
$\theta_A(t)=\theta_A(0)-E_At$.

>From this it follows that under quantum dynamics i.e
Schr\H{o}edinger equation, the motion of the system point
is {\em not over} the entire Hilbert space but over the
subspace defined by constant $|C_A(t)|^2$ for each $A$.
The motion is in fact on the $N_A$-torus spanned by
the angle variables $\theta_A$ where $N_A$ is the number
of (non-degenerate)energy eigenvalues in the interval
considered. Furthermore, the motion on this $N_A$-torus
is such that the angular velocity ${d\theta(t)\over dt}$
is constant in every direction and equal to $-E_A$. For
a macroscopic system the Hamiltonian $H$ will be
generically so complex that this motion will densely
fill the entire $N_A$-torus and because of the uniform
velocity in every direction the time spent by the system
point in any interval $[\{\theta_A\},\{\theta_A+\delta\theta_A\}]$
 is exactly proportional to the {\em volume}
of the interval. In this sense the quantum motion
is quasi-ergodic. The precise conditions to be fulfilled
by the spectrum of eigenvalues of $H$ will be discussed
later but it suffices to stress here that they are
fairly generic and unlike the classical case do not
require special assumptions about {\em chaos}.
Now consider the expectation value of an observable $O$
in the state $|\psi(t)\ran$:
\be \label{4}
\lan\psi(t)|O|\psi(t)\ran = \sum_{A,B}~C_B^{*\psi}~
C_A^ \psi~e^{-i(E_A-E_B)t}~~O_{BA}
\ee
and a time-average of this expectation value over a
duration $T$ centred at time $\tau$
\br \label{5}
\bar O(\tau) &=& {1\over 2T}~\int_{\tau -T}^{\tau +T}~~
\lan\psi(t)|A|\psi(t)\ran\nonumber\\
& =& \sum_{B,A} C_B^{*\psi}~C_A~O_{BA}~e^{-i(E_A-E_B)\tau}
{\sin {(E_A-E_B)T}\over (E_A-E_B)T}
\er
Now the main problem is that this expression has an
explicit dependence on the parameters $C_A^\psi$ of
the initial state and {\em can not} be replaced by an
ensemble average that is {\em insensitive} to the {\it
initial state}. This is the {\em mathematical 
formulation}
of the problem to be solved i.e interpret quantum
statistical mechanics in such a way that the
time-average {\em does not remember} the initial state.

It is necessary to make more precise the notion of
the time-average. Note that
\be \label{6}
{\sin(ET)\over(ET)}_{T\rightarrow\infty}\rightarrow \delta(E)
\ee
The time average in eqn (\ref{5}) then becomes
\be \label{7}
\bar O(\tau) = \sum_A |C_A^\psi|^2 O_{AA}.
\ee
At this stage nothing more can be said. If we however
consider some special class of systems it is possible
to make more statements.
For example, Berry \cite{Berry} has conjectured
that for {\em classically chaotic} systems energy
eigenfunctions behave like Gaussian random variables.
More precisely consider expanding the exact energy
eigenstates $|A\ran$ in some {\em orthonormal basis}
$|I\ran$
\be \label{8}
|A\ran = \sum_I C_A^I|I\ran
\ee
Then the Berry conjecture amounts to saying that
$C_A^I$ are {\it independently distributed random
numbers}. At this stage it is sufficient to just
specify the {\it two-point correlation} of this
distribution
\be \label{9}
\lan C_A^{*I}C_B^J\ran_{ens}= {1\over N_A}\delta_{AB}
\delta_{IJ}
\ee
It is easy to see that eqn (\ref{9}) is compatible
with unitarity. Despite the apparent basis dependence
of this criterion, it is actually {\em basis
independent}.To see this let us expand $|A\ran$ in
another orthonormal basis $I'$
\be \label{10}
|A\ran = \sum_I' C_A^{I'}|I'\ran
\ee
with
\be \label{11}
|I'\ran = B^{I'}_I|I\ran
\ee
Then
\be \label{12}
C_A^{I'} = \sum_I B_I^{I'}C_A^I
\ee
and
\br \label{13}
\lan C_A^{*I'}C_B^{I'}\ran_{ens}&=&\sum_{IJ}B_I^{*I'}B_J^{J'}\lan C_A^{*I}C_B^J\ran_{ens}\nonumber\\
&=&{1\over N_A}\sum_{IJ}B_I^{*I'}B_J^{J'}\delta_{AB}\delta_{IJ}\nonumber\\
&=&{1\over N_A}\delta_{AB}\delta_{I'J'}
\er
where we have made use of the unitarity of the transformation
matrix $B_I^{I'}$. It should be noted that this proof works only
when the bases ${I},{I'}$ etc. are not derived from the energy basis
by application of {\it fixed} unitary transformations on the
energy-eigenfunction basis. Now let us consider the
ensemble-average of eqn (\ref{7}):
\br \label{14}
\lan\sum_A |C_A^\psi|^2 O_{AA}\ran_{ens}&=&\sum_{AIJ}|C_A^\psi|^2 \lan C_A^{*I} C_A^J\ran_{ens}O_{IJ}\nonumber\\
&=&{1\over N_A}\sum_{AIJ}|C_A^\psi|^2\delta_{IJ}O_{IJ}\nonumber\\
&=&{1\over N_A}\sum_A|C_A^\psi|^2\sum_I O_{II}\nonumber\\
&=&{1\over N_A}\sum_I O_{II}
\er
Let us also consider the average of $O$ at a particular instant given
by eqn (\ref{4}):
\br \label{15}
\lan\sum_{AB}C_B^{*\psi}C_A^\psi e^{-i(E_A-E_B)t} O_{BA}\ran_{ens}
&=&\sum_{ABIJ}C_B^{*\psi}C_A^\psi \lan C_A^{*I}C_B^J\ran_{ens}O_{IJ}e^{-i(E_A-E_B)t}\nonumber\\
&=&{1\over N_A}\sum_{ABIJ}C_B^{*\psi}C_A^\psi \delta_{AB}\delta_{IJ}e^{-i(E_A-E_B)t}O_{IJ}\nonumber\\
&=&{1\over N_A}\sum_{AI}C_A^{*\psi}C_A^\psi O_{II}\nonumber\\
&=&{1\over N_A}\sum_I O_{II}
\er
which is the same as eqn (\ref{14}). Thus in these cases the time-average
is equal to the value at any given instant of time once the ensemble
average is taken
and in both of 
them all memory
of the initial state is lost. The drawback with this picture is that the
equality seems to hold at arbitrary times $t$ while one should only
expect it at late times. In other words there is no way to understand
"thermalization times" in this picture.

For eqn (\ref{6}) to hold
it actually suffices for $T >> {1\over \delta}$ where
$\delta$ is the spacing between the exact energy
eigenvalues. As this spacing typically decreases
exponentially with the size of the system the
corresponding $T$ is exponentially large in system size.
{\em This is analogous to the classical case where for
very very large times the Poincare recurrence theorem
\cite{arnold,chandra} guarantees the equality of time
and ensemble averages.}

Clearly what's important are not such astronomically large time-scales.
Instead, consider the more moderate and intermediate
time-scale\\ ${1\over \Delta}~< T~<< {1\over \delta}$:
then one has
$$
{\sin(E_m-E_n)T\over(E_m-E_n)T}\simeq 1
$$
and consequently
\be \label{16}
\bar O(\tau)=\sum C^{*\psi}_B~C^\psi_A~~e^{-i(E_A-E_B)\tau}A_{BA}
\ee
Again if the spectrum of the Hamiltonian is
sufficiently complex and there are no degeneracies, for
any reasonable value of $\tau$ there will be
cancelations among all the $A\ne B$ terms and one
again gets back to eqn (\ref{7}). Actually for such
complex systems even the average at any particular
instant(except for pathologically small values of $\tau$)
becomes the same as eqn (\ref{7}) and one does not
even need to do any time averaging. But eqn (\ref{7})
retains memory of the initial state and at this stage
it is not possible to recover the basis of statistical
mechanics unless one postulates additional assumptions
like Berry's conjecture etc.
(we will show later ways of going beyond such
restrictive assumptions; that will be the main result
of this paper).

 In  \cite{sred} there was a proposal to consider a different
approach to this problem by assuming that observables
of interest are of the type:
\be \label{17}
 O_{BA}~=~O_0~\delta_{BA}~~+R_{BA}
\ee
with $R_{BA}$ being "small".

 Now too $\lan\psi(t)|O|\psi(t)\ran$ is of the same
form as eqn(\ref{7}) except for small corrections
due to the $R_{BA}$ terms. Once again if we interpret
this restriction in the Schr\H{o}dinger picture
one would get eqn (\ref{7}) to be true at all times
which does not make much sense. A possibility is to
interpret $\lan\psi(t)|O|\psi(t)\ran$ as
$\lan\psi|O(t)|\psi\ran$ in the Heisenberg picture
and assume eqn (\ref{17}) only for late times. Even then
one has to still introduce some "ensemble average"
along the lines of Berry's conjecture and there seems
to be no scope for addressing the issue of
"thermalization time" at all. We will see later that
eqn (\ref{17}) is {\em too strong} a restriction on
macroscopic observables and in fact implies that in
the classical limit these observables are
{\em constant} on the entire energy surface.
\subsection{Earlier approaches}
Much after we had finished our work( to be described
in detail in secs 4, 5 and 6) we came to know of the
fundamental paper on this subject due to J. von Neumann
\cite{JvN} through ref \cite{GOM}. Subsequently
we have traced Pauli and Fierz's \cite{pauli} sequel
to von Neumann's work as well as Van Kampen's
formulation \cite{kampen} of the problem. There are
differences between von Neumann's and Van Kampen's
approaches but they are essentially similar in spirit
and both give a quantum mechanical equivalent of
the classical phase space cell decomposition. While
von Neumann uses fairly rigorous methods to show that
the time average of the expectation values of the
macroscopic observables (defined suitably by him)
approaches arbitrarily closely the microcanonical
average( he also shows that a suitably defined entropy,
different from what is known as von Neumann entropy
in the literature, also approaches the entropy of
the microcanonical distribution which is the quantum
version of the H-theorem). Van Kampen establishes
a master equation for the probability distribution
for finding the system in various phase cells given the
distribution at $t=0$ and argues that it has all the
properties of a Markov process whence the problem
becomes identical to the one in classical statistical
mechanics. While in von Neumann's treatment the issue
of thermalization time is not very transparent, in
Van Kampen's treatment it can be handled in principle
exactly as in classical statistical mechanics.

While the
mathematical formalism in all three approaches have
strong overlaps,
our  treatment is rather different in its
interpretation of coarse graining in quantum mechanics
which is somewhat kinematical in origin in both
von Neumann's and Van Kampen's treatments. We seek a
dynamical origin for this coarse-graining. In our approach, the
extra soft quanta not only play a passive role as the
degrees that are coarse grained away, they directly provide
the mechanism for thermalization. Thermalization takes
place due to emission and absorption of these soft quanta.
Furthermore in order for this thermalization to happen
it is crucial that they have the properties described
in Sec 2.4 and 2.5. Thus the reader will see that
the setup of Section 4,
where the solution to the problem of ergodicity is
given in an abstract way
is, perhaps not surprisingly,
mathematically identical to the setup of von Neumann. 
Both of them describe a quantum mechanical analogue of Gibbsian
coarse graining. Section 5 and Section 6 where the soft quanta
play out their role in thermalization have, no obvious
counterpart in
von Neumann's discussion.

We include a short description of the works of von Neumann and van Kampen
for  reasons of completeness and historical accuracy.
This should also provide a perspective for the whole
discussion.
This section is not a prerequisite for understanding
the rest of this paper. For convenience we have given
at appropriate places below our notation used in Section 4
corresponding to each of von Neumann's symbols.

\subsubsection{Quantum Statistical Mechanics of von Neumann}
The three essential ingredients of this approach are
i)energy-surfaces,ii)phase cells belonging to a
particular energy surface and iii)micro-states spanning
each phase cell.

{\bf Energy Surfaces:}The {\em exact} Hamiltonian of the system is taken to
be $H$ whose spectrum is assumed to be discrete
for simplicity.If $\Delta E$ is the macroscopic
resolution with which energy measurements can be made,
the energy levels are divided into groups of width
$\Delta E$ each and the groups are labeled by
$a=1,2,...$. This means that only energy levels
belonging to different groups are macroscopically
distinct.The energy eigenvalues are labeled
$W_{\rho,a}$ with $\rho = 1,2,...,S_a$.
(In Section 4 we have used $A$ to label the exact energy
eigenstates. $S_a$ corresponds to $N_A$ of Section 4, 
where we
have restricted ourselves to one energy surface.)
The eigenfunctions are labeled $\phi_{\rho,a}$. Thus
$S_a$ is the number of {\em microscopic} states
spanning the energy surface $a$. The projection operator
corresponding to $\phi_{\rho,a}$ is denoted by
${\bf P}_{\phi_{\rho,a}}$.

{\bf Macroscopic Observables:}
von Neumann considers all macroscopic observations
to be simultaneous and therefore {\em all} macroscopic
observables to be {\em mutually commuting}.
(The word ``macroscopic'' is used here in the sense
of Quantum Measurement Theory. It does {\em not}
mean macroscopic in the sense of statistical mechanics
(i.e. pressure, density, etc ).
Indeed they correspond to {\em micro-state} variables
such as the position coordinates of molecules of a gas
except that {\em macroscopic measurements} can not 
resolve their values within a phase cell.
These observables are thus a subset of  the $O$  of Section 4).
Their
{\em simultaneous} eigenfunctions are denoted by
$\omega_{\lambda,p},\lambda=1,2,..,s_p$ where distinct
values of $p$ denote distinct values of the observables
but all states with the same $p$ but different
$\lambda$ have the same values for all observables.
(In Section 4 we have used `$a$` for $\lambda$, and
`$i$` for $p$. $s_p$ is $N_a$ of Section 4. The simultaneous eigenstates
there are denoted by $| i,a \rangle$.)

The projection operator for the state
$\omega_{\lambda,p}$ is denoted by
${\bf P}_{\phi_{\lambda,p}}$ and the density matrix
signifying an equal mixture of $\omega_{\lambda,p}$
for the same $p$ but for all possible values of
$\lambda$ is denoted by
\be \label{18}
{1\over s_p} {\bf E}_p \equiv \sum^{s_p}_{\lambda}{\bf P}_{\omega_{\lambda,p}}
\ee
Eqn (\ref{18}) can be taken as the density matrix for
a "phase space cell".
In von Neumann's paper the explicit construction of
these mutually commuting macroscopic observables is
not given and the discussion not very transparent.
Van Kampen treats this in a more transparent manner.
Consider any observable $O$ and express it in the
exact energy-eigenfunction basis
\be \label{19}
O=\sum_{\rho,a,\sigma,b} O_{\rho,a;\sigma,b}|\phi_{\rho,a}\ran\lan\phi_{\sigma,b}|
\ee
Van Kampen argues that the matrix elements of the
time-dependent operator $O(t)$
\be \label{20}
O(t)_{\rho,a,\sigma,b}=O_{\rho,a,\sigma,b}e^{-i(W_{\rho,a}-W_{\sigma,b})}
\ee
are associated with {\em very rapid fluctuations}
whenever $a\ne b$.Thus it makes sense to replace the
microscopic operator $O$ by
\be \label{21}
O'=\sum_{\rho,a,\sigma,a} O_{\rho,a;\sigma,a}|\phi_{\rho,a}\ran\lan\phi_{\sigma,a}|
\ee
Equivalently
\be \label{22}
O'_{\rho,a;\sigma,b}=O_{\rho,a;\sigma,a}~\delta_{ab}
\ee
Van Kampen also introduces the coarse-grained energy
operator to be
\be \label{23}
H' = \sum_a W_a \sum_\rho^{S_a}{\bf P}_{\phi_{\rho,a}}
\ee
Note that all the energy eigenvalues belonging to
an energy-surface have been replaced by a single
value $W_a$. Now it is easy to see that $H',O'$
commute so they can be
{\em simultaneously diagonalised}:
\br \label{24}
H'|\omega_{\mu,a}\ran &=& W_a|\omega_{\mu,a}\ran\nonumber\\
O'|\omega_{\mu,a}\ran &=& A_{\mu,a}|\omega_{\mu,a}\ran
\er
He now introduces the coarse-grained observables by
grouping the eigenvalues $A_{\mu,a}$ into phase cells
in such a way that
only those belonging to different groups(cells) can be
distinguished macroscopically i.e through macroscopic
measurements.Let these groupings be labeled by
$\nu=1,2,...N_a$ ($N_a$ of von Neumann is $N_i$ of Section 4). Then
\be \label{25}
{\tilde O} = \sum_a\sum_\nu A_{\nu,a}\sum_{\lambda}^
{s_{\nu,a}}{\bf P}_{\omega_{\lambda,\nu,a}}
\ee
Now it is clear that all such macroscopic observables
commute with each other with $\omega_{\lambda,\nu,a}$
as their simultaneous eigenfunctions. Furthermore,in
this approach, the phase cells, labeled by $\nu,a$
are neatly partitioned into disjoint sets by the
energy surfaces.

It is instructive to compare this construction with
von Neumann's treatment( and later with ours). von
Neumann argues that any operator with a boolean
spectrum( eigenvalues $0$ or $1$) is a macroscopic
observable and every macroscopic observable has
a spectral decomposition of the type
\be \label{26}
A=\sum_p c_p {\bf E}_p
\ee
He then considers the function $f_a(x)$ such that $f_a=1$
 if $x$ belongs to the energy eigenvalues of the
energy surface labeled by $a$ and $0$ otherwise. Then
the operator $f_a(H)$ has eigenvalues $0,1$ and is
thus a macroscopic observable. By eqn(\ref{26}) it
must admit the decomposition
\be \label{27}
f_a(H)=\sum_p f_a^p {\bf E}_p
\ee
On the other hand
\be \label{28}
f_a(H)=\sum_{\rho=1}^{S_a}{\bf P}_{\phi_{\rho,a}}
\ee
As both $f_a(H)$ and ${\bf E}_p$ for every p are equal to
their own squares and as the product of two distinct
${\bf E}_p$'s is zero, it follows that ${f_a^p}^2 = {f_a^p}$
for every $p$ i.e $f_a^p$ is either $1$ or $0$.
Relabeling all those $p$'s for which$f_a^p=1$ by
$\nu,a$ we have
\be \label{29}
\sum_\rho^{S_a} {\bf P}_{\phi_{\rho,a}} = \sum_\nu^{N_a} {\bf E}_{\nu,a}
\ee
i.e we again have a unique partitioning of all the
phase cells among the energy surfaces. The equivalent 
of this in our approach is given by the construction in sec. 4.1.

{\bf Microcanonical ensemble:}\\
We briefly sketch JvN's proof showing the asymptotic
equivalence of the time averaged expectation value and
the microcanonical average.
We denote the operator occurring in eqn (\ref{29}) by
$\Delta_a$. It is then clear that
${1\over S_a}\Delta_a$ is the density matrix for
an equal mixture of energy eigenstates belonging
to the energy surface $a$ and as such represents
the density matrix for the microcanonical ensemble associated
with that energy surface. If the initial state $\psi$ is such that
$(\Delta_a\psi,\psi)$ is nonzero for several values of $a$, one
should consider a mixture of microcanonical ensembles for each $a$
with weight factor $(\Delta_a\psi,\psi)$(this is the probability
of finding the system in the energy surface $a$). Thus the microcanonical
density matrix in the general case is
\be \label{30}
U_\psi = \sum_{a=1}^\infty {(\Delta_a\psi,\psi)\over S_a}\Delta_a
\ee
In our treatment later on we shall consider only one energy surface.

{\bf Quantum Ergodicity}

Next JvN considers the system to be initially in a state
\be \label{3.2.31}
|\psi\ran = \sum_{a=1}^\infty\sum_{\rho=1}^{S_a} r_{\rho,a}e^{i\alpha_{\rho,a}}
|\phi_{\rho,a}\ran ~~~~(r_{\rho,a}\geq 0,0\leq\alpha_{\rho,a}< 2\pi)
\ee
then
\be \label{3.2.32}
|\psi(t)\ran = \sum_{a=1}^\infty\sum_{\rho=1}^{S_a}r_{\rho,a}e^{-i({2\pi\over h}
W_{\rho,a}t-\alpha_{\rho,a})}
\ee
Introducing the abbreviations
\be \label{3.2.33}
x_{\nu,a}=({\bf E}_{\nu,a}\psi(t),\psi(t))~~~~u_a=(\Delta_a\psi(t),\psi(t))=(\Delta_a\psi,\psi)
\ee
one considers a macroscopic observable $A$
\be \label{3.2.34}
A=\sum_{a=1}^\infty\sum_{\nu=1}^{N_a}~\eta_{\nu,a}{\bf E}_{\nu,a}
\ee
The expectation value of $A$ in the state $\psi(t)$ is then
\be \label{35}
E_A(\psi(t))=(A\psi(t),\psi(t))= \sum_{a=1}^\infty\sum_{\nu=1}^{N_a
} \eta_{\nu,a}({\bf E}_{\nu,a}\psi(t),\psi(t))
=\sum_{a=1}^\infty\sum_{\nu=N_a}\eta_{\nu,a}x_{\nu,a}
\ee
while the microcanonical average is
\br \label{36}
E_A(U_\psi)=Tr AU_\psi &=&Tr(\sum_{a=1}^\infty\sum_{\nu=1}^{N_a}\eta_{\nu,a}{\bf E}_{\nu,a}
\sum_{a'=1}^\infty\sum_{\nu'=1}^{N_{a'}}{u_{a'}\over S_{a'}}{\bf E}_{\nu',a'})\nonumber\\
&=&\sum_{a=1}^\infty\sum_{\nu=1}^{N_a}\eta_{\nu,a}{u_a s_{\nu,a}\over
 S_a}
\er
On using the eqn (\ref{36}) and Schwarz's inequality JvN finally obtains
\be \label{37}
(E_A(\psi(t))-E_A(U_\psi))^2\leq {\bar \eta}^2\sum_{a=1}^\infty\sum_{\nu=1}^{N_a}
{S_a\over s_{\nu,a}u_a}[x_{\nu,a}-{s_{\nu,a}u_a\over S_a}]^2
\ee
where
\be \label{38}
{\bar \eta}^2 = \sum_{a=1}^\infty\sum_{\nu=1}^{N_a}{s_{\nu,a}u_a\over S_a}\eta_{\nu,a}
\ee
JvN finally establishes that the time average of the lhs of eqn (\ref{37})
is bounded by ${\bar \eta}^2 max_a\{{2N_a\over S_a}\}$.
\subsubsection{Van Kampen's approach}
Van Kampen expands the initial state $\psi$ not in terms
of the exact eigenfunctions of the Hamiltonian but
in terms of the simultaneous eigenstates
$\omega_{\lambda,\nu,a}$ of the coarse grained
observables( we are retaining the notation of JvN):
\be \label{39}
|\psi\ran = \sum_{\lambda,\nu,a} b_{\lambda,\nu,a}|\omega_{\lambda,\nu,a} \ran
\ee
The probability of finding the system in the phase cell
$\nu,a$ is given by
\be \label{40}
P_{\nu,a}=\sum_{\lambda=1}^{s_{\nu,a}}|b_{\lambda,\nu,a}|^2
\ee
The time dependence of the state is given by
\be \label{41}
|\psi(t)\ran = \sum_{\lambda,\nu,a}b_{\lambda,\nu,a}(t)|\omega_{\lambda,\nu,a}\ran
\ee
where
\be \label{42}
b_{\lambda,\nu,a}(t)=\sum_{\lambda',\nu',a'}\lan \omega_{\lambda,\nu,a}|U(t)|\omega_{\lambda',\nu',a'}\ran
\ee
and $U(t)$ is the time-evolution operator. The
probability of finding the system in the phase cell $\nu,a$ at time $t$ is given by
\be \label{43}
P_{\nu,a}(t) = \sum_{\stackrel{\lambda',\nu',a'}{\lambda'',\nu'',a''}}
\sum_{\lambda}\lan\omega_{\lambda,\nu,a}|U(t)|\omega_{\lambda',\nu',a'}\ran
\lan \omega_{\lambda,\nu,a}|U(t)|\omega_{\lambda'',\nu'',a''}\ran^*
b_{\lambda',\nu',a'}(0)b^*_{\lambda'',\nu'',a''}(0)
\ee
Invoking the argument that the summation consists of many wildly fluctuating
terms and all the non-negative terms cancel i.e only surviving terms are
${\nu',a'}={\nu'',a''}$ and $\lambda'=\lambda''$.This reduces eqn(\ref{41})
to
\be \label{44}
P_{\nu,a}(t)=\sum_{\nu',a'}\sum_{\lambda'}\sum_{\lambda}
|\lan\omega_{\lambda,\nu,a}|U(t)|\omega_{\lambda',\nu',a'}\ran|^2 |b_{\lambda',\nu',a'}(0)|^2
\ee
Van Kampen further argues that in a sum of type
$$
\sum_i\alpha_i\beta_i
$$
where both $\alpha_i,\beta_i$ are {\it rapidly fluctuating} but {\it positive}
it is a good approximation to evaluate the sum as
$$
{1\over G}\sum_i\alpha_i\sum_i\beta_i
$$
With this approximation eqn (\ref{44}) becomes
\be \label{45}
P_{\nu,a}(t) = \sum_{\nu',a'}{1\over s_{\nu,a}}
\sum_{\lambda,\lambda'}
|\lan\omega_{\lambda,\nu,a}|U(t)|\omega_{\lambda',\nu',a'}\ran|^2
P_{\nu',a'}(0)
\ee
With the definition
\be \label{46}
T_t(\nu,a|\nu',a') {\stackrel {def}{=}}
{1\over s_{\nu,a}}\sum_{\lambda,\lambda'}
|\lan\omega_{\lambda,\nu,a}|U(t)|\omega_{\lambda',\nu',a'}\ran|^2
\ee
eqn (\ref{45}) can be recast as
\be \label{47}
P_{\nu,a}(t) = \sum_{\nu',a'}T_t(\nu,a|\nu',a') P_{\nu',a'}(0)
\ee
As there is nothing special about the instant $t=0$ it follows that
\be \label{48}
P_{\nu,a}(t_1+t_2) = \sum_{\nu',a'}T_{t_2}(\nu,a|\nu',a') P_{\nu',a'}(t_1)
\ee
Equivalently
\be \label{49}
T_{t_1+t_2}(\nu,a|\nu',a')=\sum_{\nu'',a''} T_{t_2}(\nu,a|\nu'',a'')T_{t_1}
(\nu'',a''|\nu',a')
\ee
Along with the obvious properties of $T_t(\nu,a|\nu',a')$
\be \label{50}
T_0(\nu,a|\nu',a')=\delta_{\nu,a;\nu',a'};~~~T_t(\nu,a|\nu',a')\geq 0;~~~
\sum_{\nu,a}T_t(\nu,a|\nu',a')=1
\ee
which follow
from the orthogonality of $|\omega_{\lambda,\nu,a}\ran,|\omega_{\mu,\nu',a'}\ran$
for distinct $\{\nu,a\},\{\nu',a'\}$ and the {\em unitarity} of $U(t)$, one
concludes that $T_t$ is a {\em Markov Process}. One knows from the
Frobenius-Perron theorem that there is always an equilibrium distribution
$P^{eq}_{\nu,a}$. From the spectrum of the operator $T(\nu,a|\nu',a')$ one
can also deduce the {\em thermalization times}.

Before concluding this section it is instructive to derive the so called
Master equation. For this one solves for $T_{\Delta t}$ as
\be \label{51}
T_{\Delta t}(\nu,a|\nu',a')=\delta_{\nu,a;\nu',a'}\{1-\Delta t\sum_{\nu'',a''}
W_{\nu'',a'';\nu,a}\}+\Delta t W_{\nu,a;\nu',a'}
\ee
Then one gets the differential form of the Chapman-Kolmogorov equation
\be \label{52}
{\partial\over \partial t}T_t(\nu,a|\nu',a')=
\sum_{\nu'',a''} W_{\nu,a;\nu'',a''}T_t(\nu'',a''|\nu',a')-W_{\nu'',a'';\nu,a}
T_t(\nu,a|\nu',a")
\ee
This can be recast in terms of $P_{\nu,a}(t)$ as
\be \label{53}
{d\over dt}P_{\nu,a}(t)=\sum_{\nu',a'}\{W_{\nu,a;\nu',a'}P_{\nu',a'}
-W_{\nu',a';\nu,a}P_{\nu,a}(t)\}
\ee
which is the {\em master equation}.
\section {Quantum coarse graining and emergence of the microcanonical
ensemble.}
\setcounter{equation}{0}

As we have stated already the major conceptual problem in Quantum statistical
 mechanics
is that of understanding how a quantum state initially described by a pure 
density
matrix can ever resemble the state of thermodynamic equilibrium which is
clearly
described by a mixed density matrix.
\subsection{Coarse graining}
As was clearly stressed by Gibbs \cite{gibbs} long ago, coarse graining is 
essential to understanding
even the most basic features of classical statistical mechanics. We now introduce
our proposal
for a {\em quantum coarse graining} and show subsequently that the microcanonical
distribution then follows for a large class of systems.

We consider two types of mutually commuting variables $I,{\cal A}$ and their
simultaneous eigenstates
$|i,a\ran $. Later on we will define $|i,a\ran $ only in the {\em subspace spanning the
states
relevant for the microcanonical ensemble}. The $i$ will eventually correspond to
the labels of the usual micro-states entering
the description of the statistical system and the precise nature of $a$ are not
specified
at present. Alternately one can think of $|i,a\ran $ as the labeling of micro-states on
the fine scale and
$|i\ran $ as the labeling of the micro-states at the coarse grained level.

It is assumed that to a very good approximation the 
${\cal A}$ variables couple
{\em weakly}
to the $I$-variables in the total Hamiltonian i.e
\be \label{4.1}
H_{tot}=H^{(i)}+H^{(a)}+\lambda~~H^{(i,a)}
\ee
with $\lambda$ very small. Let $|A^*\ran $ be the eigenstates of $H^{(i)}$ {\bf that
lie in
the range $E_0-\Delta E<E_0+\Delta$}. We then form the
matrix $I^*_{AB} = \lan B^*|I|A^*\ran $ which is $N_i\times N_i$. Now let $|i\ran $ be the
eigenstates of
$I^*$ and let $|i,a\ran= |i\ran \otimes|a\ran $ where $|a\ran $ are the eigenstates of $\cal A$
and let
us also assume that there are $N_a$ such states. (A special choice
could be $I=H^{(i)}$, in which case $|i\rangle = |A^* \rangle $.)

The important point is that the $N_i\times N_a$ states $|i,a\ran $ continue to form a
basis in terms of
which all the exact energy eigenstates in the microcanonical band can still be
expanded. It is
of course important to state that even a small perturbation characterized by
$\lambda$ can completely
alter the nature of the eigenstates of the Hamiltonian $H$ because of the exponential
crowding of the "unperturbed" states but the {\em number} of eigenstates remains
unchanged.
The parameters of the microcanonical band $E_0,\Delta$ are also most likely
affected by
the perturbation, but as we shall never need their precise values it does not matter
for the ensuing discussion.

We shall only consider those observables that are {\em insensitive} to the $a$
labels i.e $\lan i,a|O|j,b\ran
=O_{ij}\delta_{ab}$. This is in keeping with the spirit of
coarse graining described in Section 2.2.

\subsection{Some preliminaries}
Consider some arbitrary state in the subspace we have considered
\be \label{4.2}
|\psi\ran =\sum C^{\psi}_A |A\ran
\ee
where now $|A\ran $ are the {\em exact} energy eigenstates of the system.
Further, let
\be \label{4.3}
|A\ran  = \sum_{i,a}~~C_{A}^{*ia}~~|i,a\ran
\ee
with the inverse expansion
\be \label{4.4}
|i,a\ran  = \sum_A~~C^{ia}_A~~|A\ran
\ee

The coefficients $C_A^{ia}$ satisfy the following unitarity conditions:
\be \label{4.5}
\sum_A C_A^{*ia}~C_A^{i'a'}=\delta_{ii'}\delta_{aa'}~~~~~~\sum_{ia} C_A^{ia}C_{A'}^{*ia}=\delta_{AA'}
\ee

In terms of these definitions we can rewrite
\be \label{4.6}
\lan \psi|O|\psi\ran  = \sum_{A,B}~~C^{*\psi}_B~~C^{\psi}_A~~\lan B|O|A\ran
\ee
as
\be \label{4.7}
O_{\psi\psi}~~=~~\sum_{a,ij}~\sum_{A,B}~C^{*\psi}_B~~C^{\psi}_A~~
C_{A}^{*ia} ~ C_{B}^{ja} O_{ij}
\ee
Now introduce
\be \label{4.8}
\sum_a C_{A}^{ja} C_{B}^{*ka}=~~\delta_{jk}~~\delta_{AB}~~P^j_A+R^{jk}_{AB}
\ee
This is to be understood as a {\em diagonal $+$ off-diagonal} split in
 the sense
that by
definition
\be \label{4.9}
P^j_A~~=~~\sum_b C_{A}^{jb}~C_{A}^{*jb}
\ee
Consequently
\be \label{4.10}
R^{kk}_{AA}=0
\ee
(No sum on indices.)
We need to prove a few important properties of $P$ and $R$. Let $A\ne B$;
 putting $j=k$ and
summing over j one gets
\be \label{4.11}
\sum_k R^{kk}_{AB} = 0
\ee
Therefore
\be \label{4.12}
\sum_k R^{kk}_{AB}=0
\ee
always. Further,
\be \label{4.13}
\sum_k~~P^k_A~~=1  ~and ~~ also~ \sum_A P^k_A=N_a
\ee
as follow from the unitarity eqn (\ref{4.5}).  Likewise by putting $A=B$ and summing
over $A$ one gets
\be \label{4.14}
\sum_A~~R^{kj}_{AA}=0
\ee

\subsection{Emergence of the microcanonical distribution}
Let us consider a pure quantum state which at $t=0$ is given by
\be \label{4.15}
|t=0\ran ~=~|\psi\ran ~=~\sum_A~~C^{\psi}_A~~|A\ran
\ee
Now it is straightforward to show that
\be \label{4.16}
\lan \psi(t)|O|\psi(t)\ran ~~=~~\sum_{A,j}~~C^{*\psi}_A~~C^{\psi}_A~~P^j_A~~O_{jj}+
\sum_{ABjk}~~C^{*\psi}_B~C^{\psi}_A~E^{-i(E_A-E_B)t}~~R^{kj}_{AB}~O_{jk}
\ee
Separating out the $A=B$ contribution in the second term one finds
\br  \label{4.17}
\lan \psi(t)|O|\psi(t)\ran ~~=~~& &\sum_{A,j}~~C^{*\psi}_A~~C^{\psi}_A~~P^j_A~~O_{jj}
 +\sum_A~~C^{*\psi}_A~C^{\psi}_A\sum_{jk} R^{kj}_{AA}~O_{jk}\nonumber\\
&+& \sum_{A\ne B}~~\sum_{jk}C^{*\psi}_B~
C^{\psi}_A~E^{-i(E_A-E_B)t}~~R^{kj}_{AB}~O_{jk}
\er

We can further exploit the redundancy introduced by $a$ states by
defining an equivalence class of states as follows:
define the unitary operator $U^{\alpha}$ by
\be \label{4.18}
U^{\alpha}|a\ran = \sum _b U^{\alpha}_{ab}|b\ran
\ee
which is some unitary transformation on $|a\ran$ space. Clearly
it does not affect the expectation values of our observables of interest, $O$,
since $O$ is insensitive to $a$. Thus if
\be \label{4.19}
U^{\alpha}|\psi(t) \ran = |\psi ^{\alpha}(t)\ran
\ee
we have:
\be \label{4.20}
\lan \psi ^{\alpha}(t)|O|\psi ^{\alpha}(t) \ran = \lan \psi(t) |O|\psi(t) \ran
\ee
Consequently
\be \label{4.21}
\lan \psi (t) |O|\psi (t) \ran = {1\over N_{\alpha}}\sum _{\alpha}
\lan \psi ^{\alpha} (t) |O|\psi ^{\alpha}(t) \ran
\ee
Here $N_{\alpha}$ is the number of such $U$ matrices. This is up to us and we
can choose these matrices to satisfy
\be   \label{4.22}
\sum _{\alpha}U^{\alpha}_{ac} U^{* \alpha}_{bd}= \mu \delta _{ab}
\delta _{cd}
\ee
If we multiply by $\delta _{ab}$ and sum over $a,b$, we get, using the
unitarity of the matrix,
\[
\sum _\alpha \delta _{cd} = \mu N_a \delta _{cd}
\]
This gives $N_\alpha = \mu N_a$.  The conditions (\ref{4.22}) are
$N_a^4$ conditions on $N_\alpha N_a^2$ variables. Thus we need $N_\alpha = N_a^2$ at least. So we take $N_\alpha = N_a^2$ which gives $\mu = N_a$.

It should be noted that eqn (\ref{4.21}) is valid {\em only at a particular
instant $t$} as it can be valid for {\em all $t$} only if $U$ commutes
with the microscopic Hamiltonian $H$. We can think of $|\psi^{\alpha}(t)\ran$
as time evolved (after $t$) from some state $|\psi^{\alpha}\ran$. In other
words
\be \label{4.23}
|\psi^{\alpha}(t)\ran=e^{-iHt}|\psi^{\alpha}\ran
\ee
Again eqn (\ref{4.22}) is only valid for time $t$. Actually $|\psi^{\alpha}\ran$
explicitly depends on $t$ unless $[U,H]=0$.
Let
\be \label{4.24}
|\psi^{\alpha}\ran=\sum_A C_A^{\psi^{\alpha}}|A\ran
\ee
Then eqn(\ref{4.21}) implies
\be \label{4.25}
\sum_A C_A^{\psi^{\alpha}} e^{-iE_At}|A\ran=U_{\alpha}\sum_B C_B^{\psi}e^{-iE_Bt}|B\ran
\ee
Fromthis it follows that
\be \label{4.26}
C_A^{\psi^{\alpha}}=\sum_B e^{i(E_A-E_B)t}\lan A|U_{\alpha}|B\ran C_B^{\psi}
\ee
Let us evaluate $\lan A|U_{\alpha}|B\ran$ now:
\br \label{4.27}
\lan A|U_{\alpha}|B\ran &=& \sum_{ijab} C_B^{*jb}C_A^{ia}\lan ia|U_{\alpha}| jb \ran \nonumber\\
&=&\sum_{iab} C_B^{*ib} C_A^{ia} U_{\alpha}^{ba}
\er
Thus
\be \label{4.28}
C_A^{\psi^{\alpha}}=\sum_{Biaa'} e^{i(E_A-E_B)t} C_B^{*ia'} C_A^{ia} U_{\alpha}^{a'a} C_B^{\psi}
\ee
On introducing
\be \label{4.29}
|\psi(t)\ran = \sum_{ia} D_{ia}^{\psi}(t)|ia\ran
\ee
it is easy to see that
\be \label{4.30}
D_{ia}^{\psi}(t)=\sum_B C_B^{*ia} C_B^{\psi} e^{-iE_Bt}
\ee
hence
\be \label{4.31}
C_A^{\psi^{\alpha}}=\sum_{iaa'} e^{iE_At} C_A^{ia} D_{ia'}^{\psi}(t) U_{\alpha}^{a'a}
\ee

We now use all this in equation (\ref{4.21}).
\be \label{4.32}
{1\over N_{\alpha}}\sum _{\alpha}
\lan \psi ^{\alpha} (t) |O|\psi ^{\alpha}(t) \ran
= {1\over N_\alpha}\sum_{\alpha ABkle} C^{*\psi ^\alpha}_B
C^{\psi ^\alpha}_A C^{*ke}_AC^{le}_B O_{kl}e^{-i(E_A-E_B)t}
\ee
On using eqn (\ref{4.31})
\br \label{4.33}
e^{-i(E_A-E_B)t}\sum_{\alpha} C_B^{*\psi^\alpha}C_A^{\psi^\alpha}
&=&\sum_{ijacbd} D^\psi _{ic}(t)D^{*\psi}_{jd}(t)\sum _\alpha
U_\alpha ^{ca}U_{*\alpha}^{db} C_A^{ia}C^{*jb}_B\nonumber\\
&=&\mu\sum_{ijac} D^\psi _{ic}(t)D^{*\psi}_{jc}(t) C_A^{ia}C^{*ja}_B\nonumber\\
&=&\mu\sum_{ijc} D^\psi _{ic}D^{*\psi}_{jc}[P_A^i\delta_{AB}\delta_{ij}+R_{AB}^{ij}]  
\er
where we have made use of eqn(\ref{4.30}) and eqn(\ref{4.8}).
Plug this into (\ref{4.32}) to get
\be \label{4.34}
{\mu\over N_\alpha} \sum_{ijklcAB}D^\psi _{ic}(t)D^{*\psi}_{jc}(t)~[P_A^i \delta _{AB}\delta _{ij}+ R_{AB}^{ij}]
[P_A^k\delta _{AB}\delta _{kl}+ R_{BA}^{lk}]O_{kl}
\ee

We expand to get
\[
{1\over N_a}\sum_A [ D^\psi _{ic}D^{*\psi}_{ic}P_A^iP_A^kO_{kk}+
D^\psi _{ic}D^{*\psi}_{jc}R_{AA}^{ij}P_A^kO_{kk} + D^\psi _{ic}D^{*\psi}_{ic}
P_A^i R_{AA}^{kl}O_{kl}]
\]
\be \label{4.35}
+{1\over N_a}\sum_{A\ne B}D^\psi _{ic}D^{*\psi}_{jc}R_{AB}^{ij}R_{AB}^{kl}O_{kl}
\ee
Summation over all other indices as well as the time dependence of
$D^\psi_{ia}$'s is understood.

If $P^j_A$  are such that there is a {\em very weak}
dependence on either $j$ or $A$, we can draw
the following additional conclusions:
\be \label{4.37}
P^j_A~=~{1\over N_i}
\ee
This follows from eqn(\ref{4.13})

The following is one of the ways of realising eqn (\ref{4.37}): $C_A^{ia}C_A^{*ia}$
is the overlap between $|i,a\ran$ and $|A\ran$ states. If a state
$|i,a\ran$ has equal amounts of $|A\ran$ states then  $C_A^{ia}C_A^{*ia}$
will be more or less independent of $A$. This is equivalent to saying that
 the perturbation $H^{i,a}$ ``thoroughly mixes'' up the eigenstates.
Of course in
general if we plot  $C_A^{ia}C_A^{*ia}$ as a function of $A$,
while we expect it to be independent of $A$ on the average over a range,
 we also
expect the plot to be full of spikes and  dips.
The sum over $a$ smoothes out these fluctuations and should
produce a smooth constant function.
This is similar in spirit to the Berry conjecture 
though much weaker than it. 

On noting that $\sum_A R_{AA}^{ij}=0$, the final result for
$\lan\psi(t)|O|\psi(t)\ran$ is

\be \label{4.38}
\lan \psi(t)|O|\psi(t)\ran ~~~~{1\over N_i}~~\sum_j~~O_{jj}
+{1\over N_a}\sum _{i,j,k,l,c,A\ne B}D^\psi _{ic}(t)D^{*\psi}_{jc}(t)R_{AB}^{ij}R_{AB}^{kl}O_{kl}
\ee

We have used the fact that the sum over $A$ gives a factor of $N_aN_i$
in the first term.  Note that the time dependent part has an explicit
$1\over N_a$ in front.

It is imperative to show that the second term is
negligibly small compared to the first if we are to
demonstrate the emergence of the microcanonical
ensemble.
The spectrum of the Hamiltonian for a typical macro-system will be so
complex that the second term
will go to $0$ for large times. Then we have our desired result.

It is also important to know how fast the second term
vanishes and for this purpose some estimation of the
magnitude of the second term is important.

When the expansion coefficients $C_A^{ia}$ are
identically distributed independent random variables
such an estimation can be carried out quite accurately.
In accordance with general ideas
expressed in \cite{Berry,sred}, such a
circumstance could arise when the quantum system is
{\em classically chaotic}.
The nature of the eigenstates then
is such that at high enough energies, all eigenstates 
look more or less the same. But it should be emphasised that our 
approach does not really require any strong
statement like this but only the weaker statement of eqn (\ref{4.38}).

The assumption about wildly fluctuating phases
allows us to estimate the magnitude of $R_{AB}^{ij}$. The magnitude
of any given $C_A^{ia}$ can be estimated to be
$1\over \sqrt {N_iN_a}$.
Thus $R_{AB}^{ij}$ is a sum of $N_a$ terms each of magnitude
$1\over N_i N_a$ and random phase. Thus we expect it to be $O({{\sqrt N_a }
\over N_a N_i})\approx {1\over N_i \sqrt N_a }$. Also $P_A^i \approx {1\over
N_i}$.  The off diagonal part  thus has an extra factor of
$1\over \sqrt N_a$ relative to the diagonal part.

We can similarly estimate the off-diagonal part in (\ref{4.38}).
We have
\[
\lan \psi(t)| O | \psi (t)\ran =
\sum _{j,k.A,B,b} C_A^{*\psi}C_B^{\psi}C_A^{jb}C_B^{*kb}e^{-i(E_B-E_A)t}O_{jk}
\]

The off diagonal part is when $A \ne B$. There are $\approx N_A^2$ such terms,
each having magnitude $1\over {N_A^2}$. Using the random phase approximation
and including the sum over $j,k,b$ we get
$ \approx {1\over N_A^2}N_A N_i \sqrt N_a = {1\over \sqrt N_a}$.
This is as expected since the averaging over $N_a$ fine grained
micro-states is expected to produce just such a suppression. Indeed,
this was the motivation for introducing the $a$ 
variables. This also agrees with v. Neumann's estimates
on noting that his ${N_a\over S_a}$ is our $N_a$.

The time-dependent term is nevertheless very important
for determining the
{\em thermalization time}. Thus we have the result that to a very good
accuracy (of order $1\over N_a^{1/2}$), the effective density
matrix for large times
is given by
\be \label{4.39}
\rho~~= {1\over N_i}\sum_i~|i\ran \lan i|~~=~~{1\over N_A}~~\sum_A~~|A\ran
\lan A|
\ee

It should be noted that the emergence of the microcanonical distribution is
happening for
non-trivial systems. This can be seen by examining 
eqns (\ref{4.16}-\ref{4.17}) at $t=0$ where
 there is no
trace of the microcanonical distribution anywhere.

In section 6 we will calculate the leading order time dependent
correction in a simple example that confirms the above expectations.

One should also ask what happens if we start off the system in an exact
energy eiegenstate. In that case we get for the expectation value
of $O$
\be \label{4.40}
\lan \psi (t) | O \psi (t) \ran = \sum _i p_A ^i O_{ii} +
\sum _{ij}R_{AA}^{ij}O_{ij}
\ee

Note that if $O$ is a reasonable observable we should find
that $\lan i| O^2|i\ran - \lan i|O|i\ran^2 < \lan i|O|i\ran ^2$.
This means that $\sum _j O_{ij}O_{ji} \approx O_{ii}^2$. Thus
one expects ${O_{ii}\over O_{ij}} \approx \sqrt N_j$ if we assume
random phases or ${O_{ii}\over O_{ij}} \approx  N_j$ if they add in phase.
If we now take into account this relative magnitudes
of $O_{ij}$ versus $O_{ii}$, one gets that the off diagonal term is suppressed
by a factor between $\sqrt{N_aN_i}$ and ${\sqrt {N_a}} N_i$. If $N_a$ is
sufficiently large this could mean that for such systems even
if the system starts of in an exact energy eigenstate it can be described
by the microcanonical distribution. 

\section{Soft Quanta and the Resonance Picture}
\setcounter{equation}{00}

In this section we will attempt to describe the general
phenomenon of thermalisation in slightly different terms. Instead
of working with the full Hamiltonian of the $i$ and $a$ system
and the exact eigenstates, let us consider the same system from the point
of view of time dependent perturbation theory. We think of the
$i$ system as an atom with discrete states and the $a$ system as
a collection of harmonic oscillators describing long wavelength
photons.

As a first step let us treat the electro-magnetic field
classically and let us think of the atom as a two level system. We are
familiar with this ``NMR'' type of situation. When the electro-magnetic
RF-pulse is the right frequency, we have resonant absorption
and emission of photons. As we apply the RF pulse we expect the
two level system to oscillate between the two states at the Bohr frequency
and on the average there is equal probability to be in either
of the two states. This happens even when the coupling is very weak.
In terms of time independent perturbation theory (in the NMR case,
 we can go to the rotating frame and make the problem
time independent) the off diagonal terms are very large, close to
 resonance,
compared to the energy splitting. Thus the energy eigenstates have
equal amounts of up and down spin states.

This physical intuition can be applied to our problem. First
of all the electro-magnetic field has to be treated quantum mechanically.
Thus there is spontaneous emission in addition to induced emission
and absorption. If the strength of the electro-magnetic field
is large then this can be neglected. In oscillator language,
the field oscillators should be excited to a high enough quantum number.

Secondly, and more importantly we have a continuum of oscillators,
both for emission and absorption. Here again we can apply our
intuition from atomic physics. The effect of the continuum on the
discrete state is to introduce an anti-Hermitian term in the effective
Hamiltonian that gives the discrete states an imaginary ``width''.
This makes the time evolution non-unitary and gives the states
a lifetime by making the wave function decay. Similar mathematical
manipulations can be done for the traced density matrix calculation to
show within a perturbative approach that the off diagonal terms decay with time
and the diagonal terms tend
to become equal. The decay of the off diagonal terms was explained
in Section 3 as being due to random phase cancelation. 
Here we
give arguments for seeing this more explicitly. Just as
in the usual atomic physics calculation, the
excited state decays by emission of energy in the form of a photon, similarly
the off diagonal terms also decay as the photon is emitted and
absorbed repeatedly. Just as the photon escapes
in an apparently irreversible way into the continuum,
never to return
(in the limit of the Poincare recurrence time $\rightarrow \infty$),
 so do the correlations escape irreversibly into the
continuum. \footnote{In the AdS/CFT correspondence, this continuum
of soft photons represents the far infrared of the CFT. In the
bulk, this would be the region in the interior of the black hole.
The irreversibility described above is the irreversibility of the black
hole horizon. This was discussed in \cite{KRBS}}.

This language is also suitable for explaining intuitively why the
traced matrix has equal diagonal terms. When the number
density of soft quanta is large, then the probability,
$P_{ij}$, of transition from $i$ to $j$ is equal to the reverse.
We expect that the change in the $i$'th diagonal term in the
density matrix is proportional to $P_j P_{ji}-P_i P_{ij}$.
When $P_{ij}=P_{ji}$, this will stop changing only when
$P_i=P_j$. This does not
constitute a proof that the system will go toward equilibrium in
general. But it allows us to intuitively understand a criterion
for the emergence of microcanonical distribution. The main requirements are thus
two:

i) There should be a continuum of soft excitations with energies
containing the range of $\delta _i $ to $\Delta E$ in order for
resonant absorption and emission to take place.
In particular if $\delta _i > \Delta E$ there will not be
any ergodization.

ii)The number of these should be large. We expect this number to be
proportional to $E_{\gamma}\over {\hbar \omega}$, where $E_\gamma$
is the total energy in soft quanta. Since $\hbar \omega \le \Delta E$,
we have $E_\gamma >> \Delta E$. Also one expects $E >> E_\gamma$.
Thus $E >> \Delta E$, as expected.

If we have a large number of
hard photons as well, i.e. those with frequency larger than
$\Delta E$, one would have to include them in the
$i$-system. Otherwise the microcanonical description would not
be a good approximation.

While the discussion in Section 4 is completely general, it
is somewhat abstract.
The picture
in this section is more intuitive. We can use this picture
to attempt an answer to the question of when, in a given
system, one can expect self thermalization. The difficult
part presumably, is to identify the $i$ and $a$ variables
unambiguously.
\section{Two Discrete States Coupled to a Continuum}
\setcounter{equation}{00}

The $|i\rangle \otimes |a\rangle$ system can be thought of as being composed
of an atom coupled to the electro-magnetic field. This is denoted schematically
in Figure 1 below. What is shown are the energy levels of the uncoupled
system or equivalently, the situation where the coupling constant is
set to zero. When the interaction is turned on we expect the energy
levels of the atom to get broadened.

\begin{figure}[htbp]
\begin{center}
\epsfig{file=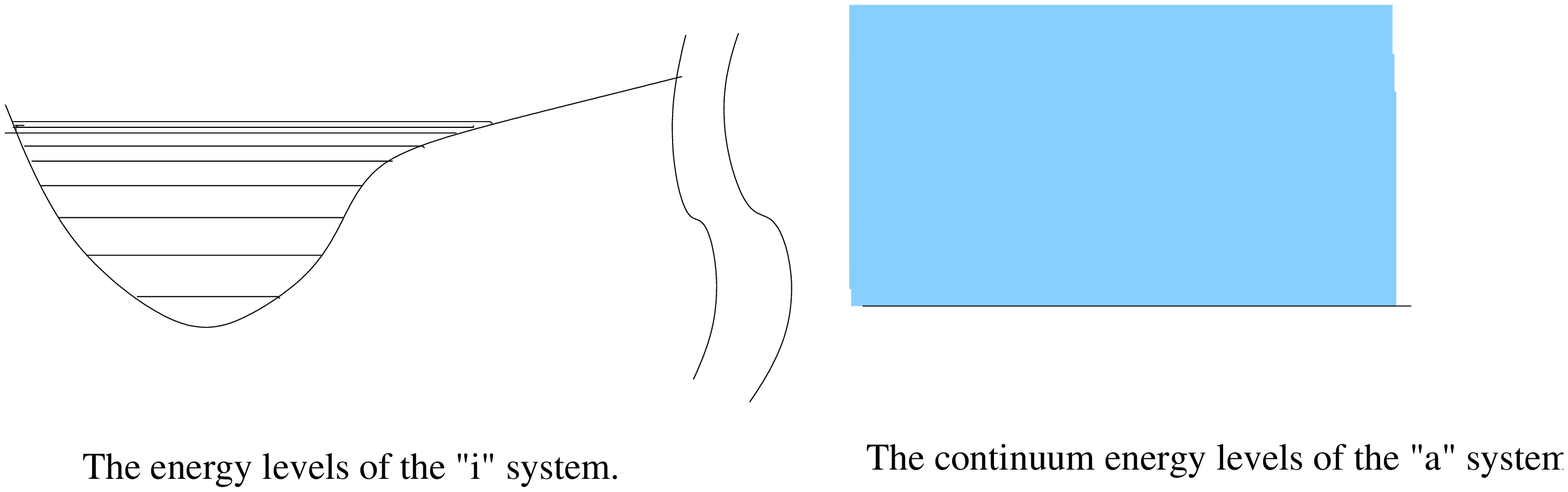, width= 12 cm,angle=0}
\vspace{ .2 in }
\begin{caption}
{Discrete states coupled to a continuum}
\end{caption}
\end{center}
\label{fig2}
\end{figure}

\subsection{Relaxation of Density Matrix - Thermalization}

We make the approximation namely that the $i$ system has only
two states, $|0\rangle $ and $|1\rangle$. 
Furthermore we will assume that the continuum degrees are those
of a continuum
of harmonic oscillators. Harmonic oscillators
are easy to analyze and also representative of the real
situation where we expect $a$ degrees to be soft quanta. 
The frequencies
of the harmonic oscillators are assumed to start at zero and form a continuum.
In practice we just need the discrete spacing to be very small
compared to the (inverse) time scales of interest. So if $\omega$ denotes 
the frequency of the oscillators,
${1\over \delta E_a} = {1\over \delta \omega} >> T$. On the other hand
$T$ should be large enough that the system has time to thermalize.
In the previous section we saw that this time scale was set by $\gamma$.
Thus $\gamma T > 1$ and so $\delta \omega << \gamma$. $\gamma$
in turn is fixed by the matrix elements $\langle 1,0|V|0,a\rangle$.
Thus here also we require that the off diagonal terms generated by
 the interaction with the $a$ degrees of freedom must not be too small -
this is expected whenever one has a resonance situation , i.e.
$\hbar \omega = \Delta E$.

To do the calculation we will use the technique of influence
functionals due to Feynman and Hibbs \cite{feyn}.

Following \cite{feyn} we let 
$q$ be the coordinates of the $i$ system 
\footnote{Though the $q$-system is actually a two level
system, we have adopted a notation wherein the $q$-variables look continuous. The path integrals over $q$ are
to be understood as appropriate matrix elements in the
two level system.}
and $Q$ that of the $a$ system. If the system starts
off at time $t=0$ in the state $\Psi (q_i,Q_i,0)$, then the
density matrix at time $t=T$ is given by (The subscript $i$ stands for
`initial', and $f$ for `final')

\[
\rho (q_f,q'_f,Q_f,Q'_f;T)=
\int ...\int dq_i dq'_i dQ_i dQ'_i
{\cal D} q(t){\cal D}q'(t) {\cal D}Q(t) {\cal D} Q'(t)
\]
\be
e^{i[S(q)-S(q')] + i[S(Q)-S(Q')] +i[S_{int}(Q,q)-S_{int}(Q',q')]}
\Psi (q_i, Q_i,0)\Psi ^* (q'_i,Q'_i,0)
\ee

with boundary conditions:
\[
q(0)=q_i , q(T)=q_f , Q(0)=Q_i, Q(T)=Q_f
\]
\be
q'(0)=q'_i , q'(T)=q'_f , Q'(0)=Q'_i, Q'(T)=Q'_f
\ee
The action $S(q)$ for the $i$-system will not be specified. However we will
assume, as stated before that there are only two discrete states
$|0\rangle , |1 \rangle$. We will further make the simplifying 
assumption that $\langle 0 | q |0\rangle =\langle 1 | q |1\rangle =0$
and that $\langle 0 | q |1\rangle \ne 0$.

The $a$-system is a harmonic oscillator. Actually it is an infinite number
of harmonic oscillators, so in fact there should be an infinite number
of $Q$ variables. For transparency of presentation 
we first consider just one of them. As they are
independent the generalization to arbitrary numbers
or even a continuum is easy.

The wave function, $\Psi (q_i,Q_i,0)$, is assumed to factorize as
\be
\Psi (q_i,Q_i,0)= \phi _I (q_i)\chi _a (Q_i)
\ee

Here
\be
|\phi _I\rangle = \alpha |0\rangle + \beta |1 \rangle
\ee
with  $ |\alpha |^2 + |\beta |^2 =1$. Let us first consider the case
\be
\chi (Q) \approx e^{-{Q^2\over 2 \omega}}
\ee

which is the ground state of a harmonic oscillator of frequency $\omega$.  Furthermore
we take $S_{int}(q,Q) \approx C\int dt q(t)Q(t)$.

We are interested in the density matrix traced over the $Q$ variables:
\[
\int dQ_f \rho (q_f,q'_f,Q_f,Q_f,T) \equiv \rho _Q(q_f,q_f',T)
\]
If we imagine doing all the $Q,Q'$ integrals we get an expression of the form
\be   \label{rho}
\rho _Q(q_f,q'_f,T)= \int ...\int dq_i dq'_i {\cal D} q(t){\cal D}q'(t)
e^{i[S(q)-S(q')]}\phi _I(q_i)\phi ^*_I(q'_i)F(q,q').
\ee

Here $F(q,q')$ is the result of doing all the $Q$ integrals and incorporates
the entire effect of the $Q$-system, called 
{\em environment} by Feynman and Hibbs, on the 
$q$-system. This is the influence functional.
For the special case of quadratic $S(Q)$ and bilinear $S_{int}(q,Q)$ it has been
shown in \cite{feyn} that it has the form,
\be   \label{IF1}
F(q,q')=e^{-\int \int dt dt' [q(t)-q'(t)][\alpha (t,t')q(t')-\alpha ^* (t,t')q'(t')]}
\ee

For a harmonic oscillator of frequency $\omega $,
in its {\em ground state}
we have $\alpha (t,t')= \alpha (t-t')$ and
also
\be   \label{alpha}
\int d\tau e^{-i\nu \tau }\alpha (\tau )
\equiv a(\nu ) \approx {C\over \omega ^2}\delta (\nu +\omega )
\ee
This is what Feynman and Hibbs call a
{\it cold-environment} and for such a situations
transitions pumping energy into the system are unlikely.

When we have a continuum of oscillators we simply have to integrate this
expression (\ref{alpha}) over $\omega$.
We thus insert (\ref{IF1}) into (\ref{rho}) and evaluate it perturbatively.  

\subsubsection{Zeroth Order}

\br
\rho _{00}= \int dq_f dq_f' \phi _0^*(q_f)\phi _0 (q_f')\rho_Q(q_f,q'_f,T)\nonumber \\
\rho _{10}= \int dq_f dq_f' \phi _1^*(q_f)\phi _0 (q_f')\rho_Q(q_f,q'_f,T)
\er
and so on.
Thus
\be
\lan 0 |\rho |0\ran = 
\lan 0 | e^{-iHT}(\alpha |0\ran + \beta |1\ran )(\lan 0|\alpha ^* + \lan 1|\beta ^* )
e^{iHT}|0\ran = |\alpha |^2
\ee
\be
\lan 1 |\rho |1\ran = |\beta |^2
\ee
\be
\lan 1 |\rho |0\ran = 
\lan 1 | e^{-iHT}(\alpha |0\ran + \beta |1\ran )(\lan 0|\alpha ^* + \lan 1|=
\beta ^* )
e^{iHT}|0\ran = \alpha ^* \beta e^{-i(E_1-E_0)T}
\ee
\be
\lan 0 |\rho |1\ran = \alpha \beta ^* e^{+i(E_1-E_0)T}
\ee

\subsubsection{First Order}
 We now go to the next order. We bring down one power of $\alpha (t,t')$.

We write it as the sum of four terms:
\[
\delta \rho _{Qj0}(T)=
-\int \int dq_f dq'_f \phi ^* _j(q_f)\phi _0 (q_f') \int {\cal D}q(t){\cal =
D}q'(t)
\int dq_i \int dq'_i
\]
\be   \label{drho}
e^{i[S(q)-S(q')]}[a+b+c+d]\phi _I (q_i)\phi _I^*(q'_i)
\ee

where the subscript $j$ on $\phi$ can be  $0,1$ and
\br
a &=& \int dt \int dt' q(t)q(t')\alpha (t,t') \nonumber \\
b &=& -\int dt \int dt' q'(t)q(t')\alpha (t,t') \nonumber \\
c &=& -\int dt \int dt' q(t)q'(t')\alpha ^*(t,t') \nonumber \\
d &=& \int dt \int dt' q'(t)q(t')\alpha ^*(t,t')
\er

The calculations are done in the appendix. Here we will simply give the result.

We get for the first order correction to $\rho _{00}$ the following:
\be   \label{31}
\delta \rho _{Q00}(T)= -T\{ |\alpha |^2 [2a_R(\Delta E) + 2 a_R(-\Delta E)]-
2a_R(-\Delta E) \}|\lan 0 |q|1\ran |^2
\ee
 where $\Delta E = E_1-E_0$, and

 Let us first assume that $\Delta E >0$. Using (\ref{alpha}), which says that
$a_R(\Delta E )=0$, we find
\be   \label{32}
\delta \rho_{Q00}(T) = -T[(|\alpha |^2 -1)2a_R(-\Delta E )]|\lan 0|q|1\ran |^2
\ee
This goes to zero when $|\alpha | \rightarrow 1$.

If $\Delta E <0$ we get
\be   \label{33}
\delta \rho _{Q00}(T)= -T |\alpha |^2 2 a_R(\Delta E )|\lan 0|q|1\ran |^2
\ee

When $E_1 > E_0$ one expects the system to decay to its lower energy state
$|0\ran$ with the emission of a soft quanta. The decay stops when the
probability of occupation of the lower energy state is one. Also as (\ref{32})
shows $\delta \rho$ is increasing with time as it should be.
Thus one expects $\beta (T) \approx \beta (0) e^{-\gamma T}$.  So
$|\alpha (T)|^2 = 1- |\beta (T)|^2 \approx |\alpha (0)| ^2 +|\beta (0)|^2
2\gamma T$.  Thus $\delta \rho \approx |\beta (0)|^2 2\gamma T$. This agrees
with (\ref{32}) for $\gamma = a_R(-\Delta E)|\lan 0|q|1\ran |^2$.

Similarly when $E_0>E_1$, the probability of occupation of $E_0$
decreases with time as $|\alpha (T)|^2 \approx |\alpha (0)|^2 e^{-2\gamma T}
$ and so $\delta \rho \approx -|\alpha (0)|^2 2\gamma T$ as
shown by (\ref{33}).

Similarly we can look at the off diagonal terms
\[
\delta \rho_{Q10} = -T \alpha ^* \beta e^{-i(E_1-E_0)T}|\lan 0|q|1\ran |^2
[a(-\Delta E) + a(\Delta E)^*]+
\]
\be   \label{34}
[\int _0^Td\tau e^{-i(E_1-E_0)\tau}] \alpha \beta ^* e^{-i(E_1-E_0)T}
(\lan 1 |q |0\ran )^2 [a(0) + a(0)^*]
\ee

If $\Delta E \ne 0$, the integral over $\tau$ in the second term of
(\ref{34}) is vanishingly small relative to the first term. We are left with
\be \label{6.20}
\delta {\rho}_{Q10} = -T \alpha ^*\beta e^{-i(E_1-E_0)T}|\lan 0|q|1\ran |^2
[a(\Delta E)^* + a(-\Delta E )]
\ee
This is also as expected because the product $\alpha (T) \beta (T)$
decays as $e^{-\gamma T}$ (one of the terms decays and the other goes
to 1).

While all this is
as expected, this does not really demonstrate the De-coherence that we
are after. In order to demonstrate decoherence we must show that
the off diagonal terms decay even when the diagonal terms do not.
In order to obtain a situation where the probabilities of occupation
of the states $|0\ran ,|1\ran$, are equal, one must have the external
oscillators in excited states as well so that the system can absorb soft
quanta.
An easy route would be to let the two states be exactly
degenerate so that energy conservation allows transitions
in either direction. But in such a situation it is always
possible to find orthonormal linear combinations of the two
states in which the the operator $q$ is {\it diagonal}. It is
then a simple exercise to show that in this basis the diagonal
elements of the density matrix do not change at all. Even on
physical grounds exact degeneracy is unreasonable to assume.
What is more reasonable is to consider a situation
in which $\Delta E\ne 0$ but "small" compared to some
relevant energy scale
in the problem. Nevertheless, we need to calculate influence
functionals for situations where the system can both absorb
and give energy to the quantum environment. The influence
functional calculated in Feynman and Hibbs for the case where
all the environment oscillators are in the ground state will
not suffice. While the generic case of the environment
oscillators in arbitrary excited states is too difficult to
handle, the influence function for the case where the
environment states are {\it coherent states} or {\it number states}
can be
explicitly evaluated and this suffices for our purpose.

A partial form of this result is already available in Feynman
and Hibbs (see eqn 8.141). They considered the amplitude
$f(b,a)$
that a {\it forced} harmonic oscillator goes from state
$f$ to state $g$ where
\be
f(x)~=~({M\omega\over \pi\hbar})^{1\over 4}
~e^{-({M\omega\over \pi\hbar})(x-a)^2}~~~~~~~~
g(x)~=~({M\omega\over \pi\hbar})^{1\over 4}
~e^{-({M\omega\over \pi\hbar})(x-b)^2}
\ee
Their explicit expression for $f(b,a)$ is
\be
f(b,a)~=~A~e^{[-{M\omega\over 4\hbar}(a^2+b^2-2ab~e^{-i\omega T})+
i({M\omega\over 2\hbar})^{1\over 2}
(a\eta+b\eta^*e^{-i\omega T})]}
\ee
where
\be
A~=~e^{-i\omega T/2}
~e^{
-{1\over 2M\omega\hbar} \int_0^T~\int_0^t~\gamma(t)\gamma(s)
e^{-i\omega(t-s)}
}
\ee
Of course, the states $f(x),g(x)$ are {\em coherent states}
of the harmonic oscillator but with {\em real} parameters.
It suffices, for our purposes, to generalise this result to the case where $f(x)$ is a coherent state with {\it complex}
parameter $\xi$ while leaving $b$ still real. The
corresponding results  are
\be
f(x)~=~({M\omega\over \pi\hbar})^{1\over 4}
~e^{{\xi^2\over 2}-{|\xi|^2\over 2}}
~e^{-({M\omega\over \pi\hbar})(x-\sqrt{{2\hbar\over M\omega}}~\xi)^2}~~~~~~~~
\ee
where we have followed the Pancharatnam convention for phases.
\be
f(b,\xi)~=~{\cal A}~e^{[-{M\omega\over 4\hbar}(
{2\hbar\over M\omega}
\xi^2+b^2-2
\sqrt{{2\hbar\over M\omega}}
\xi b~e^{-i\omega T})+
i({M\omega\over 2\hbar})^{1\over 2}
(
\sqrt{{2\hbar\over M\omega}}
\xi\eta+b\eta^*e^{-i\omega T})]}
\ee
where
\be
{\cal A}~=~e^{{\xi^2\over 2}-{|\xi|^2\over 2}}
~e^{-i\omega T/2}
~e^{
-{1\over 2M\omega\hbar} \int_0^T~\int_0^t~\gamma(t)\gamma(s)
e^{-i\omega(t-s)}
}
\ee
Now it is easily seen that $G_{0\xi},G_{m\xi}$
where
$G_{fi}$ is the amplitude that the oscillator initially
in state $i$ is found in state $f$ at time T, are given
by
\be
G_{0\xi}=~e^{-{|\xi|^2\over 2}+i\xi\eta}
~e^{
-{1\over 2M\omega\hbar} \int_0^T~\int_0^t~\gamma(t)\gamma(s)
e^{-i\omega(t-s)}
}
\ee
and
\be
G_{m\xi}=~{1\over {\sqrt{m!}}}~G_{0\xi}~(\xi+i\eta^*)^m
\ee
Finally the influence functional is given by
\be
F[q,q^{\prime}]~=~\sum_m~G_{m\xi}{G^{\prime}_{m\xi}}^*
\ee
The final result can be expressed as
\be
F_{\xi}[q,q^{\prime}]~~e^{i\xi(\eta-\eta^{\prime})+i\xi^*(\eta^*-{\eta^{\prime}}^*)}
~F_0[q,q^{\prime}]
\ee
where $F_0$ is the influence functional for the case
where the environment oscillators are in the ground
state.

Thus the environmental coherent states produce an
additional influence functional whose exponent is a
linear functional of $q$. However, to the degree of
accuracy needed for our perturbative calculations
we can translate this into an effective quadratic
functional by expanding the additional term to second
order i.e
$$
-{1\over 4M^2\omega}[\xi^2~(\eta-\eta^{\prime})^2
+{\xi^*}^2~(\eta^*-{\eta^{\prime}}^8)^2
+2|\xi|^2~|(\eta-\eta^{\prime})|^2]
$$
It is sufficient to look at terms of the type
$$
\int_0^T~dt\int_0^t~ds \Delta a(t,s) q(t) q(s)
$$
to determine the {\em one complex function} $\Delta a(t,s)$
that determines the additional effective influence functional.
After some algebra it follows that
\be
\Delta a(t,s)~=~{C^2\over 4M^2\omega}~
[2\xi^2 e^{-i\omega(t+s)}+2{\xi^*}^2 e^{i\omega(t+s)}+
2|\xi|^2 (e^{i\omega (t-s)}+e^{-i\omega(t-s)}]
\ee
Unlike the case of the environment oscillators in the
ground state the complex function is not a function of
$t-s$ alone but it turns out that the parts of $a(t,s)$
that are functions of $t+s$ do not contribute to the
transition rates as the integrations over $s,t$ produce
mutually exclusive delta functions and one is
effectively left with
\be \label{6.32}
\Delta a_{eff}(t,s)~=~{C^2\over 4M^2\omega}~
2|\xi|^2 [(e^{i\omega (t-s)}+e^{-i\omega(t-s)}]
\ee
Thus when the environmental oscillators are in  coherent
states we have a mixture of a {\it cold-environment}
whose strength is {\em independent} of the coherent
state parameter $\xi$ and a {\it classical noise}
environment with $a_R(\nu)=a_R(-\nu)$. Further $a_R(\nu)$ is
 directly proportional to the level of excitation
of the coherent states. It should be
emphasised that any environment can be represented
as a mixture of a cold environment and a classical
noise environment. But what is special about the
coherent state case is that the cold component is
{\em independent} of the level of excitation of the coherent states.
This will be crucial for us later on.

In the number state case, there are no $t+s$ terms to begin
with (as shown on the Appendix) and one obtains (\ref{C28})
directly - this is (\ref{6.32}) with $|\xi |^2 $
replaced by $m$, the excitation number of the oscillator.

If we use (\ref{C28}) (or (\ref{6.32})), with large $m$
or $|\xi |^2$, we can neglect the cold term in the
influence functional. In this case $a_R(\Delta E)=
a_R (-\Delta E)$, and both are non-zero. Then we
see from (\ref{6.20}) that
\[
\delta \rho _{Q00} = -T (2 |\alpha | ^2 -1) |\lan 0|q|1\ran|^2 2a_R(\Delta E)
\]
\[
\delta \rho _{Q01}= -T \underbrace
{\alpha ^* \beta e^{-i\Delta E T}}_{\rho _{Q01}}|\lan 0 |q|1\ran|^2
2a_R(\Delta E)
\]
\be \label{6.33}
\approx ~\rho _{Q01}(e^{-\gamma T}-1)
\ee
Thus in this approximation
even when $|\alpha |^2={1\over2}$, $\delta \rho _{Q01}\ne 0$.
Equation (\ref{6.33}), if our extrapolation from the linear term
to the full exponential is correct, demonstrates decoherence. To actually
 prove decoherence
one would have get a more complete solution. Presumably this can be done numerically
in some cases.

Thus we have evidence for  decoherence in an explicit calculation in
this simplified model. The crucial point here is the appearance
of the imaginary term due to ``absorption of a soft quanta by the
environment''. We have put the quotation marks because this is really
just a convenient trick. We have made a division into two sets of variables
$i$ and $a$ with the requirement that $a$ should have a continuous spectrum
We called these ``soft quanta''. Then we showed that the same mathematical
manipulations that give rise to an imaginary part to the energy of a
discrete state can be used here to show that the off-diagonal terms of the 
density matrix decay.

We should also check that the conditions set forth at the end
of the last section are indeed satisfied. $\delta _a$ is zero for this
calculation.
Due to resonance, the off-diagonal terms are large.
Thus both conditions
are satisfied. While we
have not {\em proved} even in a non-rigourous way,
the existence of the phenomenon of self thermalization,
the calculations done here do make plausible the physical ideas of section
2 and also buttress the intuitive random phase arguments
about the off diagonal terms of the density matrix
decaying in time, that were made in section 3.

\section{Conclusions}

In this paper we have attempted to provide a physical picture
of the process of self thermalization, by which a pure state
can appear thermal if studied with coarse resolution. This phenomenon
underlies the validity of the microcanonical ensemble in quantum statistical
mechanics. It also has applications in the black hole information
paradox.

We have presented an intuitive
physical picture along with a mathematical formulation
of the process. The principal new ingredient was the introduction
of dynamical ``soft quanta'' that are not included in the
microstate descriptions. They are
coarse grained away and
in the process, remove correlations between the (coarse grained) microstates.
Thus they not only provide the quantum analogue of the Gibbsian coarse
graining of microstates (Section 4), they also thermalize the system
by resonant absorpton and emission (Section 5,6). Mathematically
this is identical to the appaearance of an imaginary part in the
energy of an excited state when it couples to a continuum.
This also provides a  way to calculate the thermalization rate.
A first order perturbation calculation (Section 6) supports this picture.  

It would be interesting to apply this picture to some interesting
systems in a quantitative way. This would be a test of the correctness
of these ideas.

\noindent
{\bf Acknowledgements}\\ We would like to thank Balram
Rai and
R. Shankar for discussions.

\appendix

\renewcommand{\thesection}{\Alph{section}}
\renewcommand{\theequation}{\thesection.\arabic{equation}}

\section{Appendix: Density Matrix Calculation}
\label{appena}
\setcounter{equation}{0}
In this appendix we derive (\ref{31}) and (\ref{34}) starting from (\ref{drho}).

We will first work out in detail the contribution of the perturbation labelled
`$a$'.

The $q$ and $q'$ integrals can be done separately. Let us denote by
$A_j$ the $q$ integral and by $A'$ ($A''$ does not depend on $j$) the $q'$ integral.
Thus
\[
A_j=\int dq_f \phi ^*_j (q_f){\cal D}q e^{iS(q)}\int dq_iq(t)q(t')\phi _I(q_i)
\]
\[
A'= \int dq'_f \phi _0(q'_f){\cal D}q'(t) e^{-iS(q')}dq'_i \phi _I^*(q'_i)
\]
\be
\delta \rho _{Qj0}(T) = \int _0^T dt\int _0^T dt' \alpha (t,t') A_j A'
\ee

{$\bf A_j:$}

\[
A_j=\int dq_f \int dq_i \phi ^* _j (q_f)\int {\cal D}q e^{iS(q)}q(t)q(t')
\phi _I(q_i)
\]
\[
 =
\sum _{m=0,1}
\lan j|e^{-iH(T-t)}q|m\ran e^{-iH(t-t')}\lan m|qe^{-iHt'}
(\alpha |0\ran +\beta |1\ran )  
\]

If $j=0$ we get
\[
A_0 =
\lan 0 |e^{-i(E_0(T-t)} q | 1 \ran e^{-iE_1(t-t')}\lan 1 | qe^{-iE_0t'} \alpha |0\ran
\]
\be
=e^{-iE_0T}e^{-i(E_1-E_0)(t-t')}\lan 0 |q|1\ran \lan 1 | q | 0 \ran \alpha
\ee

If $j=1$ we get

\[
A_1 = \lan 1 | e^{-iE_1(T-t)}q | 0\ran e^{-iE_0(t-t')}\lan 0| qe^{-iE_1t'}\beta |1\ran
\]
\be
= e^{-iE_1T}e^{-i(E_0-E_1)(t-t')}\lan 1|q |0 \ran \lan 0 | q | 1 \ran \beta
\ee

{$\bf A':$}

\[
A'=[ \int dq'_f {\cal D} q'(t) dq'_i \phi _I (q'_i)e^{iS(q')}\phi _i^*(q'_f)]^*
\]
\be
=
[\lan 0 | e^{-iHT}(\alpha | 0 \ran + \beta | 1 \ran ) ] ^*
=\alpha ^* e^{iE_0T}
\ee

Combining all the above we get the contribution due to the perturbation
$a$.
:  

\[
\delta \rho _{Q00}(T)= |\alpha |^2 \int _0^Tdt \int _0^T
dt' e^{-i(E_1-E_0)(t-t')}\alpha (t-t')
|\lan 1 |q|0\ran |^2
\]
\be
= T |\alpha |^2 |\lan 1 |q|0\ran |^2 a(E_1-E_0)
\ee

and

\[
\delta \rho _{Q10}(T) = \alpha ^* \beta e^{-i(E_1-E_0)T}\int _0^T dt
\int _0^T dt'
e^{-i(E_0-E_1)(t-t')} \alpha (t-t') |\lan 1 |q|0\ran |^2
\]
\be
=
T \alpha ^* \beta |\lan 1 |q|0\ran |^2 a(E_0-E_1)e^{-i(E_1-E_0)T}
\ee

We can similarly calculate the contribution from the other ones.

\noindent
{$\bf B_j:$}

\[
B_0= \lan 0|e^{-iH(T-t')}q|m\ran e^{-iE_mt'}\lan m
 |(\alpha |0\ran + \beta |1\ran )
\]
\be
= \beta e^{-iE_0T}e^{-i(E_1-E_0)t'}\lan 0 | q| 1 \ran
\ee

\[
B_1=\lan 1|e^{-iH(T-t')}q|m\ran e^{-iE_mt'}\lan m | (\alpha
| 0 \ran + \beta | 1 \ran )
\]
\be
= \alpha e^{-iE_1T}e^{-i(E_0-E_1)t'}\lan 1 | q|0 \ran
\ee
$\bf B':$

\[
B'=[ \int dq'_f \phi _i^*(q'_f){\cal D} q'(t) dq'_i e^{iS(q')}\phi _I (q'_i)]^*
\]
\[
= [\lan 0 | e^{-iH(T-t)}q'|m\ran e^{-iE_m t}\lan m|
(\alpha |0 \ran + \beta | 1 \ran )]^* 
\]
\be
= \beta ^* e^{+iE_0T}\lan 1 | q | 0 \ran e^{-i(E_0-E_1)t}
\ee

Combining the above

\be
\delta \rho _{Q00}= T|\beta |^2 |\lan 0 |q|1\ran |^2 a(E_0-E_1)
\ee
and
\be
\delta \rho _{Q10}=[\int _0^T d\tau
e^{-i(E_0-E_1)\tau}]
\alpha \beta ^*e^{-i(E_1-E_0)T}(\lan 1 | q | 0 \ran )^2  a(0)
\ee

(Note that the factor inside the square brackets replaces the factor
$T$ that appeared in previous terms.)

$\bf C_j:$

\[
C_0 = \lan 0| e^{-iH(T-t)}q|m\ran e^{-iE_mt}\lan m | (\alpha |0\ran + \beta
|1 \ran )
\]
\be
= \beta e^{-iE_0 T}e^{-i(E_1-E_0)t}\lan 0 | q | 1\ran
\ee

\[
C_1= \lan 1 | e^{-iH(T-t)}q|m\ran e^{-iE_mt}\lan m | (\alpha |0\ran + \beta
|1 \ran )
\]
\be
=\alpha e^{-iE_1 T}e^{-i(E_0-E_1)t}\lan 1| q | 0\ran
\ee

$\bf C':$

\[
C'= [ \int dq'_f \phi _i^*(q'_f){\cal D} q'(t') dq'_i e^{iS(q')}
\phi _I (q'_i)]^*
\]
\[
= [\lan 0 | e^{-iH(T-t')}q'|m\ran e^{-iE_m t'}\lan m|
(\alpha |0 \ran + \beta | 1 \ran )]^* 
\]
\be
=\beta ^*  e^{+iE_0 T}e^{-i(E_0-E_1)t'}\lan 1| q | 0\ran
\ee

Combining:
\be
\delta \rho _{Q00}= T |\beta |^2  |\lan 0 | q | 1 \ran |^2 [a(E_0-E_1)]^*
\ee

\be
\delta \rho _{Q10}=[\int _0^T d\tau
e^{-i(E_0-E_1)\tau}]
\alpha \beta ^*e^{-i(E_1-E_0)T}(\lan 1 | q | 0 \ran )^2  a^*(0)
\ee

$\bf D_j:$

\be
D_0= \lan 0|e^{-iHT}(\alpha |0\ran + \beta
|1 \ran )
=\alpha e^{-iE_0T}
\ee
\be
D_1= \lan 1|e^{-iHT}(\alpha |0\ran + \beta
|1 \ran )
=\beta e^{-iE_1T}
\ee

$\bf D':$
\[
D'=
[ \int dq'_f \phi _i^*(q'_f){\cal D}q'(t) q'(t') dq'_i e^{iS(q')}
\phi _I (q'_i)]^*
\]
\[
= [\lan 0 | e^{-iH(T-t)}q'|m\ran e^{-iE_m (t-t')}\lan m|q e^{-iHt'}
(\alpha |0 \ran + \beta | 1 \ran )]^* 
\]
\be
= \alpha ^* e^{iE_0T}e^{-i(E_0-E_1)(t-t')}|\lan 0|q |1\ran |^2
\ee

Combining:
\be
\delta \rho _{Q00}= T |\alpha |^2  |\lan 0 | q | 1 \ran |^2 [a(E_1-E_0)]^*
\ee

\be
\delta \rho_{Q10}=
T \alpha ^* \beta |\lan 1 |q|0\ran |^2 [a(E_1-E_0)]^*e^{-i(E_1-E_0)T}
\ee

If we add up all the contributions
$a+b+c+d$ we get the expressions (\ref{31}) and
(\ref{34}).

\section{Appendix: Influence Functional Calculation}
\label{appenb}
\setcounter{equation}{0}

In this Appendix we will derive the influence functional for the
cases when the "external" oscillators are either in coherent
states or in  number eigenstates.

We start with the kernel $K(Q_f,Q_i,Cq)$ for a forced
harmonic oscillator forced with $f(t)=Cq(t)$ given by
\be
K= \underbrace {{\sqrt {M\omega \over 2\pi i \hbar sin ~ \omega T }}}_N e^{{i\over\hbar}
S_{cl}(Q_f,Q_i,Cq)}
\ee
As shown in \cite{feyn} $S_{cl}$ is given by,
\br
S_{cl}&=& {M\omega\over 2 sin ~ \omega T}[cos ~\omega T (Q_f^2+Q_i^2)
-2Q_fQ_i\nonumber\\
&+&
{2Q_fC\over M\omega} \int _0^T q(t) sin ~ \omega t ~dt +
{2Q_iC\over M\omega} \int _0^T q(t)sin ~ \omega (T-t)~dt\nonumber\\
&-&{C^2\over M^2\omega^2} \int _0^T\int _0^t q(t)q(s)sin ~\omega (T-t) 
sin~ \omega s ~ds ~dt ]
\er

We need to calculate the following quantity:
\be \label{IF}
\int dQ_f dQ_i dQ'_i |N|^2 e^{{i\over\hbar}[S_{cl}(Q_f,Q_i,Cq) -S_{cl}(Q_f,Q'_i,Cq')}
\psi (Q_i) \psi ^* (Q_i')
\ee
The $Q$ oscillator starts off at time $t=0$ in a state $\psi$. The density matrix at time $T$ is then traced over the coordinates of
the $Q$ oscillator.  For $\psi (Q_i)$ we use the coherent states.
Thus we use
\be
f(x)~=~({M\omega\over 2\hbar})^{1\over 4}
~e^{{\xi^2\over 2}-{|\xi|^2\over 2}}
~e^{-({M\omega\over 2\hbar})(x-\sqrt{{2\hbar\over M\omega}}~\xi)^2}~~~~~~~~
\ee
where we have followed the Pancharatnam convention for phases.
It should be noted that eqn (B.3) straightaway gives the
influence functional. The method followed in \cite{feyn} is
needlessly circuitous. After obtaining the {\em nonperturbative}
expression for the coherent case influence functional which
will be shown to have a part {\it linear} in $q(t),q^{\prime}(t)$,
we will introduce an effective influence functional quadratic
in $q(t),q^{\prime}(t)$ that is obtained perturbatively. We will
obtain expressions for this effective influence functional
both for the {\em coherent states} as well as for the {\em number states}.

The $Q$-dependent terms in the integrand in (\ref{IF}) can be written as
\be
e^{-{1\over 2} [A Q_f^2 + BQ_i^2 + C{Q'}_i^2 +2DQ_fQ_i +2EQ_fQ'_i ]
+ FQ_f + GQ_i + HQ'_i }
\ee
with
\be
A=0, ~~B=-{M\omega\over \hbar \sin {\omega T}}ie^{i\omega T}, ~~C=B^*,
\ee
\be
D= {iM\omega\over \hbar\sin {\omega T}}~~,~~
E=D^*
\ee
\be
F= {iC\over \hbar\sin { \omega T}} \int _0^T (q-q') \sin {\omega t} ~dt,
\ee
\be
G= {iC\over \hbar\sin { \omega T}} \int _0^T q sin ~ \omega (T-t) ~
dt~+~\sqrt{{2M\omega\over\hbar}}\xi
\ee
\be
H= {-iC\over \hbar\sin { \omega T}} \int _0^T q^{\prime} sin ~ \omega (T-t) ~
dt~+~\sqrt{{2M\omega\over\hbar}}\xi^*
\ee

The result of doing the Gaussian integral is
\be
\sqrt{{(2\pi)^3\over 2}({\hbar\over M\omega})^3 sin^2 ~\omega T}~~
e^{{\hbar\over 4M\omega}[F^2 + (G+H)^2 + e^{-i\omega T}2FG + e^{i\omega T}2FH]}
\ee
The overall normalisation factor is
\br
{\cal N}&=&|N|^2~
\sqrt{{(2\pi)^3\over 2}({\hbar\over M\omega})^3 sin^2 ~\omega T}~~
({M\omega\over \pi\hbar})^{1/2}~
e^{\xi^2/2+{\xi^*}^2/2-|\xi|^2}\nonumber\\
&=&
e^{\xi^2/2+{\xi^*}^2/2-|\xi|^2}
\er
 Thus the final result is
\br
&~&{\cal N} exp \{{\hbar\over 4M\omega}[F^2 + (G+H)^2 + e^{-i\omega T}2FG +
 e^{i\omega T}2FH] \nonumber\\
&-&{i\over \hbar\sin{\omega T}}{C^2\over M\omega} \int _0^T\int _0^t q(t)q(s)sin ~\omega (T-t)
sin ~\omega s ~ds ~dt \nonumber\\
&+&{i\over \hbar\sin{\omega T}}{C^2\over M\omega} \int _0^T\int _0^t q'(t)q'(s)sin ~\omega (T-t)
sin ~\omega s ~ds ~dt \}
\er

As shown in \cite{feyn} the influence functional for this class
of problems has the general form
\be
F(q,q^{\prime})=\int _0^T \int _0^t [q(t)-q'(t)][a(t,s)q(s) - a ^*(t,s)q'(s)]
~dt~ds
\ee
Thus we can extract $a(t,s)$ from the $q(t)q(s)$ terms.
Calculationally it is easier to look at the $q(t)q^{\prime}(s)$
terms to get $a(t,s)$ as the terms in $S_{cl}$ that are quadratic in $q(t)$ do not contribute. We find for the $\alpha$
independent part of the $q(t)q(s)$ term the following sum of
four terms(an overall factor of ${C^2\over M\omega}$ has
been suppressed for the moment):

\[
-{1\over2 sin ^2 ~\omega T} \int _0^T \int _0^t q(t)q(s)sin ~\omega t
sin ~\omega s
\]
\[
-{1\over2 sin ^2 ~\omega T} \int _0^T \int _0^t q(t)q(s)sin ~\omega (T-t)
sin ~\omega (T-s)
\]
\[
-{e^{\i\omega T}\over 2sin^2 ~\omega T}\int _0^T \int _0^t q(t)q(s)
[sin~\omega t sin ~\omega (T-s) +sin~\omega s sin ~\omega (T-t) ]
\]
\be
-{i\over sin ~\omega T}\int _0^T \int _0^t q(t)q(s)sin~\omega (T-t)
sin ~\omega s
\ee

These can be rewritten as functions of $t-s$ and $t+s$.
The $t-s$ terms are(after restoring all factors)
\br
&-&{C^2\over M\omega}{1\over 4 sin ^2 ~\omega T}[2cos ~\omega (t-s)
-{e^{-i\omega T}} cos ~\omega (T-s+t)\nonumber\\
&-&{e^{-i\omega T}} cos ~\omega (T-t+s)
-2i\sin{\omega T}cos ~\omega (T-t+s)]
\er
They add up to
\be \label{GS}
-{C^2\over M\omega}[{cos ~\omega (t-s)\over 2} + i {sin ~\omega (t-s)\over 2}]
= -{C^2\over 2M\omega}e^{i\omega (t-s)}
\ee
This is the influence functional when the $Q$-oscillator is in
its ground state \cite{feyn}.
We also have to check that the $t+s$ terms
are zero
The $t+s$ terms are
\br
&~&{1\over 4 sin ^2~ \omega T}cos ~\omega (t+s) +
{1\over 4 sin ^2~ \omega T}cos ~\omega (2T-t-s)\nonumber\\
&~&{ie^{-i\omega T}\over 2 sin^2 ~ \omega T }cos~ \omega (T-s-t)-
{i\over 2sin \omega T}cos \omega (T-t-s)
\er

They can be seen to add up to zero.

Now we turn to the $\xi$-dependent terms.First let
us look at the terms quadratic in $\xi$:
\[
{\hbar\over 4M\omega}[({2M\omega\over\hbar})(\xi+\xi^*)^2]={1\over 2}(\xi+\xi^*)^2
\]
the exponential of which exactly cancels the factor ${\cal N}$.

\be \label{ES}
\ee

Now we look at the terms linear in $\xi$:
\[
{\hbar\over 4M\omega}[2F\sqrt{{2M\omega\over\hbar}}
(\xi e^{-i\omega T}+\xi^* e^{i\omega T})+
2\sqrt{{2M\omega\over\hbar}}(\xi+\xi^*)
{iC\over\hbar \sin{\omega T}}\int_0^T \sin{\omega(T-t)}
(q-q^{\prime})]
\]
On using the expression for $F$ and some rearrangement
it is easy to see that this equals
\be
{iC\over\hbar}\sqrt{{\hbar\over 2M\omega}}\int_0^Tdt~
(q(t)-q^{\prime}(t))
(\xi e^{-i\omega T}+\xi^* e^{i\omega T})
\ee
which can be reexpressed as
\be
iC\xi(\eta -\eta^{\prime})+iC\xi^*(\eta^*-{\eta^{\prime}}^*)
\ee
where
\be \label{eta}
\eta = \sqrt{{1\over 2M\omega\hbar}}\int_0^Tdt~q(t)e^{-i\omega t}
\ee
We can thus express the influence functional for the
case when the $Q$-oscillator is in the coherent state
as
\be
F_{\xi}(q,q^{\prime})=F_0(q,q^{\prime})~e^
{iC\xi(\eta -\eta^{\prime})+iC\xi^*(\eta^*-{\eta^{\prime}}^*)}
\ee
with $\eta$ given by eqn (\ref{eta}).
Within perturbation
theory we can expand the exponential of the linear
functional in $q,q^{\prime}$.It turns out that the
linear term does not contribute and the quadratic
terms are:
$$
-{1\over 4M^2\omega}[\xi^2~(\eta-\eta^{\prime})^2
+{\xi^*}^2~(\eta^*-{\eta^{\prime}}^8)^2
+2|\xi|^2~|(\eta-\eta^{\prime})|^2]
$$
It is sufficient to look at terms of the type
$$
\int_0^T~dt\int_0^t~ds {\Delta a(t,s)}_{coh} q(t) q(s)
$$
to determine the {\em one complex function}
${\Delta a(t,s)}_{coh}$
that determines the additional effective influence functional.
After some algebra it follows that
\be
{\Delta a(t,s)}_{coh}~=~{C^2\over 4M^2\omega}~
[2\xi^2 e^{-i\omega(t+s)}+2{\xi^*}^2 e^{i\omega(t+s)}+
2|\xi|^2 (e^{i\omega (t-s)}+e^{-i\omega(t-s)}]
\ee
Now we turn to the evaluation of the effective influence
functional when the $Q$-oscillators are in the $m$-th
number state $|m\ran $. For this we recall the
expansion of the coherent state $|\xi\ran$ in terms
of the number states:
\be
|\xi\ran=e^{-|\xi|^2/2}\sum_n~{\xi^n\over \sqrt{n!}}|n\ran
\ee
Thus
\be
F_{\xi}(q,q^{\prime})=e^{-|\xi|^2}\sum_{n,m}
{\xi^n\over\sqrt{n!}}{{\xi^*}^m\over \sqrt{m!}}~
F_{m,n}(q,q^{\prime})
\ee
Therefore to get the influence functional(in the
perturbative sense described above) for the case in
which the $Q$-oscillators are in the state $|m\ran$
one has to expand $e^{|\xi|^2}~F_{\xi}$ in
powers  of $|\xi|^2$(and quadratic in $q(t)$
and pick the coefficient of
$|\xi|^{2m}$.The $m$-th term of the expansion is
\be
{|\xi|^{2m}\over m!}~a_0(t,s)
+
{|\xi|^{2(m-1)}\over (m-1)!}~ C^2 |\eta|^2
\ee
It is elementary to show that
\be \label{C28}
a_m(t,s)=a_0(t,s)+m{C^2\over M\omega}(e^{i\omega(t-s)}
+e^{-i\omega(t-s)})
\ee


\begin{thebibliography}{999}


\bibitem{KRBS} S. Kalyana Rama and B. Sathiapalan,
{\em On the Role of Chaos in the AdS/CFT Connection}
Int. J. Mod. Phys. A14 (1999) 2635, hep-th/9905219.

\bibitem{maldacena} J. Maldacena, Adv. Theor. Math. Phys.
{\bf 2}(1998), hep-th/9711200.

\bibitem{witten} E. Witten, Adv. Theor. Math. Phys.
{\bf 2}(1998), hep-th/9802150.

\bibitem{reichl} L.E. Reichl, "A Modern Course in 
Statistical Physics", University of Texas Press, 
Austin, p.241.

\bibitem{Sinai} Ya. G. Sinai in the {\it The Boltzmann
Equation},ed. E.G.D. Cohen and W. Thirring(Springer-
Verlag, Vienna, 1973).

\bibitem{ehren} Paul and Tatiana Ehrenfest,{\it The
Conceptual Foundations of the Statistical Approach
in Mechanics},Cornell University Press,Ithaca,New
York,1959.

\bibitem{gibbs}J.W. Gibbs,"Elementary Principles in
Statistical Mechanics", Dover Edition, 1960.
\bibitem{arnold}V.I. Arnold, "Mathematical Foundations
of Classical Mechanics", Springer-Verlag,1980, p.71.

\bibitem{chandra}S. Chandrasekhar, Rev. Mod. Phys. {\bf 15} 1(1943).

\bibitem{JvN}J.v. Neumann, Zeitschrift fur Physik,
{\bf 57}, 30, 1929.

\bibitem{pauli}Sommerfeld Festschrift(Leipzig 1928)
p.30; W. Pauli and M. Fierz, Zeits. Phys.{\bf 106}
572{1937}.

\bibitem{kampen}N.G. Van Kampen {\it Fundamental 
Problems in Statistical Mechanics of Irreversible
Processes} in {\it Fundamental Problems in Statistical
Mechanics}, ed.E.G.D. Cohen,North-Holland Publishing Company,1962.

\bibitem{Biro}T. S. Biro, S. G. Matinyan and B. Mueller,
``Chaos and Gauge Field Theory'', (1994),World Scientific, Singapore ;
 T. S. Biro, C. Gong and B. Mueller, Phys. Rev. D52 (1995) 1260,
hep-ph/940932.

\bibitem{Muel} B. Mueller and A. Trayanov, Phys. Rev. Lett. 68 (1992) 3387 ;
T.S.Biro, M. Feurstein, H. Markum, hep-lat/9711002.

\bibitem{F1}
V.V.Flambaum and F.M.Izrailev,
{\em Time dependence of occupation numbers and thermalization
time in closed chaotic many-body systems},
quant-ph/0108109.

\bibitem{FI2}
V.V.Flambaum and F.M.Izrailev,
{\em Entropy production and wave packet dynamics in
the Fock space of closed chaotic many-body systems},
quant-ph/0103129.

\bibitem{I}
F.M.Izrailev,
{\em Quantum Chaos and Thermalization for Interacting Particles},
lectures given in the CXLIII Course "New Directions
in Quantum Chaos" on the International School of
Physics "Enrico Fermi"; Varenna, Italy, July 1999;
to be published in Proceedings;
cond-mat/9911297.

\bibitem{CIK}
Doron Cohen, Felix M. Izrailev, and Tsampikos Kottos,
{\em Wavepacket dynamics in energy space, RMT and
quantum-classical correspondence},
Phys. Rev. Lett. {\bf 84} (2000) 2052;
chao-dyn/9909015.

\bibitem{FGGP}
V. V. Flambaum, A. A. Gribakina, G. F. Gribakin, and
I. V. Ponomarev,
{\em Interaction-Driven Equilibrium and Statistical Laws
in Small Systems. The Cerium Atom},
Phys. Rev. {\bf E 57} (1998) 4933;
cond-mat/9711213.

\bibitem{BGIC}
F.Borgonovi, I.Guarneri, F.M.Izrailev, and G.Casati,
{\em Chaos and Thermalization in a Dynamical Model of
Two Interacting Particles},
chao-dyn/9711005.

\bibitem{FI3}
V.V. Flambaum and F.M. Izrailev,
{\em Distribution of occupation numbers in finite
Fermi-systems and role of interaction in chaos
and thermalization},
Phys. Rev. {\bf E 55} (1997) R13; 
cond-mat/9610178.

\bibitem{GOM}
Jochen Gemmer, Alexander Otte, and Guenter Mahler,
{\em Quantum Approach to a Derivation of the Second
Law of Thermodynamics},
quant-ph/0101140.

\bibitem{S}
D. L. Shepelyansky,
{\em Quantum Chaos \& Quantum Computers},
Physica Scripta {\bf T90} (2001) 112;
quant-ph/0006073. 

\bibitem{GS}
B. Georgeot and D. L. Shepelyansky,
{\em Emergence of Quantum Chaos in Quantum Computer Core
and How to Manage It},
quant-ph/0005015.

\bibitem{JS}
Ph. Jacquod and D. L. Shepelyansky,
{\em Emergence of quantum chaos in finite interacting
Fermi systems},
cond-mat/9706040.

\bibitem{LMI1}
G. A. Luna-Acosta and J. A. Mendez-Bermudez, F. M. Izrailev,
{\em Periodic Chaotic Billiards: Quantum-Classical
Correspondence in Energy Space},
Phys. Rev. {\bf E 64} (2001) 036206;
cond-mat/0105108. 

\bibitem{LMI2}
G. A. Luna-Acosta, J. A. Mendez-Bermudez, and F. M. Izrailev,
{\em Quantum-classical correspondence for local density of
states and eigenfunctions of a chaotic periodic billiard},
Phys. Lett. {\bf A 274} (2000) 192-199;
nlin.CD/0002044.

\bibitem{MS}
B. Mehlig and M. Santer,
{\em Universal eigenvector statistics in a quantum
scattering ensemble},
cond-mat/0012025.

\bibitem{P}
Don N. Page,
{\em Average Entropy of a Subsystem},
Phys.Rev.Lett. {\bf 71} (1993) 1291;
gr-qc/9305007.

\bibitem{SS}
Siddhartha Sen,
{\em Average Entropy of a Subsystem},
Phys.Rev.Lett. {\bf 77} (1996) 1;
hep-th/9601132.

\bibitem{sred}
Mark Srednicki,
{\em The approach to thermal equilibrium in quantized
chaotic systems},
J. Phys. {\bf A 32} (1999) 1163, 
cond-mat/9809360;
{\em Thermal Fluctuations in Quantized Chaotic Systems},
J.Phys. {\bf A 29} (1996) L75,
chao-dyn/9511001;
{\em Does Quantum Chaos Explain Quantum Statistical Mechanics?},
cond-mat/9410046;
{\em Quantum Chaos and Statistical Mechanics},
talk given at the Conference on Fundamental Problems
in Quantum Theory, Baltimore,
cond-mat/9406056;
{\em Chaos and Quantum Thermalization},
Phys. Rev. {\bf E 50} (1994) 888,
cond-mat/9403051.

\bibitem{Berry}
M. V. Berry,
J. Phys. {\bf A 10} (1977) 2083;
In Les Houches XXXVI,
{\em Chaotic Behaviour of Deterministic Systems},
Edited by G. Iooss, R. H. G. Helleman, and R. Stora,
North-Holland, Amsterdam, 1983;
In Les Houches LII,
{\em Chaos and Quantum Physics},
Edited by M. -J. Giannoni, A. Voros, and J. Zinn-Justin,
North-Holland, Amsterdam, 1991.

\bibitem{Sinha}
Supurna Sinha,
{\em Decoherence at Absolute Zero}, 
Phys. Lett. A 228 (1997) 1. 


\bibitem{G}
M. C. Gutzwiller,
{\em Chaos in Classical and Quantum Mechanics},
Springer-Verlag (1990).

\bibitem{HS}
H. J. Stockmann,
{\em Quantum Chaos, an Introduction},
Cambridge University Press (1999).



\bibitem{Haake}
F. Haake,
{\em Quantum Signatures of Chaos},
Springer Series in Synergetics, 54,
Springer-Verlag (2000), Second edition.

\bibitem{feyn}
R. P. Feynman and A.R. Hibbs,"Quantum Mechanics and 
Path Integrals",McGraw-Hill Publishers,1965.

\end{thebibliography}
\end{document}